\documentclass[aps,twocolumn,superscriptaddress,floatfix,longbibliography]{revtex4-1}
\usepackage{tikz}
\usepackage{lipsum}
\usepackage{graphicx}
\usepackage[export]{adjustbox}
\usepackage{amsmath,amssymb,wasysym}
\usepackage{braket}
\usepackage{mathtools}
\usepackage{mathrsfs}
\usepackage{verbatim}
\usepackage{makecell}
\usepackage{cases}
\usepackage{bbm}
\usepackage{enumitem}

\usepackage{hyperref}

\newcommand{\lmax}{\lambda_{\max}}
\newcommand{\im}{\mathbf{i}}

\newcommand{\C}{\mathbb{C}}
\newcommand{\sech}{\mathrm{sech}}

\newcommand{\sgn}{\mathrm{sign}}

\renewcommand{\Re}{\mathrm{Re}}
\renewcommand{\Im}{\mathrm{Im}}

\newcommand{\BigO}{\mathcal{O}}

\begin{document}
	\title{Scrambling versus relaxation in Fermi and non-Fermi Liquids}
	
	\author{Jaewon Kim}
	\author{Xiangyu Cao}
	\author{Ehud Altman}
	\affiliation{Department of Physics, University of California, Berkeley, CA 94720, USA}
	
	\begin{abstract}
		We compute the Lyapunov exponent characterizing quantum scrambling  in a family of generalized Sachdev-Ye-Kitaev models, which can be tuned between  different low temperature states from Fermi liquids, through non-Fermi liquids to fast scramblers. The analytic calculation, controlled by a small coupling constant and large $N$, allows us to clarify the relations between the quasi-particle relaxation rate $1/\tau$ and the Lyapunov exponent $\lambda_L$ characterizing scrambling. In the Fermi liquid states we find that the  quasi-particle relaxation rate dictates the Lyapunov exponent. In non-Fermi liquids, where $1/\tau \gg T$, we find that $\lambda_L$ is always $T$-linear with a prefactor that is independent of the coupling constant in the limit of weak coupling. Instead it is determined by a scaling exponent that characterizes the relaxation rate. $\lambda_L$ approaches the general upper bound $2\pi T$ at the transition to a fast scrambling state. Finally in a marginal Fermi liquid state the exponent is linear in temperature with a prefactor that vanishes as a non analytic function $\sim g \ln (1/g)$ of the coupling constant $g$.
	\end{abstract}
	\maketitle

	\section{Introduction}

	Thermalization takes place in isolated many-body systems because of the scrambling of information, whereby simple observables become complex and inaccessible over time. This is the mechanism that produces effective dissipation and allows relaxation dynamics to proceed. But while scrambling enables relaxation, these are distinct phenomena characterized by markedly different objects. 
	Relaxation dynamics is represented by standard retarded correlation functions and characterized by relaxation times $\tau$. 
	The dependence of $\tau$ on temperature or wave-vector can reveal the universal long time dynamics of the system.
	Scrambling, on the other hand, is described by the intermediate time growth of out-of-time order correlators (OTOC) or squared commutator~\cite{larkin,Maldacena:2015waa}:
	\begin{equation}\label{eq:OTOC_intro}
	C(t) = \left< [V(t), W(0)]^\dagger [V(t), W(0)] \right> \,.
	\end{equation}
	In a semi-classical limit, this correlation function can be interpreted as the sensitivity to initial conditions; thus it is expected to grow exponentially with time in chaotic systems, with a growth rate $\lambda_L$ called the Lyapunov exponent. While the Lyapunov exponent may not be well defined in generic quantum systems, a clear exponential growth is obtained in systems with a large number of particle flavors (large-$N$). It is natural then to ask if the Lyapunov exponent is determined by universal physics and explore its relation to relaxation time scales. These issues are perhaps best understood in the context of ``fast scramblers'', which are systems that saturate the general low-temperature bound on the Lyapunov exponent~\cite{Maldacena:2015waa}:
	\begin{equation}
	\lambda_L \le 2\pi T \,,
	\end{equation} 
	in the natural units $\hbar = 1$, $k_b = 1$, which we shall use from now on.  A prominent example of a fast scrambler is the Sachdev-Ye-Kitaev (SYK) model~\cite{Sachdev:1992fk,george-parcollet,Kitaev:2015,syk}. In general, fast scrambling occurs as a consequence of emergent conformal/reparametrization symmetry of the low-energy theory ~\cite{Kitaev:2015,syk,kitaev:soft}. 
	At the same time, the universal aspects of scrambling in other systems are not well understood. While $\lambda_L$ has been calculated in many situations, see e.g.~\cite{Patel1844,sumilan,guo,diffusive,Galitski,PhysRevD.96.065005}, few general principles have emerged. For example, it is not clear how $\lambda_L$ depends on low-energy properties such as quasiparticle decay rate, or on universal exponents in quantum critical systems.
	
	\begingroup
	\setlength{\tabcolsep}{4pt} 
	\renewcommand{\arraystretch}{1.4} 
	\begin{table}[]
		\centering
		\begin{tabular}{|c|c|c|}
			\hline
			&  quasiparticle decay $1/\tau$ & $\lambda_L$  \\
			\hline
			Fermi liquid  & $\sim g T^{1+\eta}$, $\eta > 0$  &$  \sim g T^{1+\eta}$  \\
			marginal FL  & $\sim g T \ln (1/T)$ & $ \sim g \ln (1/g) T$ \\
			non-Fermi liquid    &  $ \sim g T^{1+\eta}$, $\eta < 0$   & $ = C_\eta T  $ \\
			Fast scrambler & No quasiparticles & $2\pi T$ \\ 
			\hline
		\end{tabular}
		\caption{Temperature ($T$) dependence of quasiparticle lifetime $\tau$ and Lyapunov exponent $\lambda_L$ in different states of matter. In Fermi liquids, $\lambda_L$ is much small than $T$, and is linear in the interaction strength $g$. In non-Fermi liquids, $\lambda_L$ is $T$-linear, and the prefactor only depends on an exponent, and not on the coupling strength. Marginal Fermi liquids are distinguished by subtle log corrections. Fast scramblers have no quasiparticles and saturate the bound on scrambling.}
		\label{tab:summary}
	\end{table}
	\endgroup
	
	In this paper we shed light on these issues by calculating the Lyapunov exponent in a family of models that are analytically tractable and can interpolate continuously between slow and fast scrambling. 
	Specifically we consider the low-rank SYK model, where a large number of fermions and bosons interact with random Yukawa couplings ~\cite{masaki16,Bi:2017yvx,wang,esterlis,masaki16,Kim:2019lwh}:
	\begin{equation}
	H = \sum_{n=1}^{R} \left[ \frac12 \phi_n^2 + \sqrt{\lambda_n} \phi_n \sum_{i,j=1}^N \im u_{ij}^{(n)}\gamma^i\gamma^j  \right]      \,.
	\label{eq:lowrankH_intro}
	\end{equation}
	Here $\gamma^1,\ldots,\gamma^N$ are Majorana fermions and $\phi_1,\ldots,\phi_R$ are real bosons. The $u^{(n)}_{ij}$ are Gaussian random variables with zero mean and variance $1/N^2$ and $\lambda_1,\ldots,\lambda_R$ are $\mathcal{O}(1)$ constants that tend to a smooth distribution in the thermodynamic limit
	\begin{equation}
	\rho(\lambda) \equiv \frac1N \sum_{n = 1}^R \delta(\lambda-\lambda_n)\,.
	\end{equation}
	By tuning the distribution of the coupling strengths, it is possible to realize a number of low-energy states,  which include, besides fast scramblers, a family of Fermi liquids, non-Fermi liquids and a marginal Fermi liquid. The fast scramblers were analyzed previously~\cite{Kim:2019lwh}, and this paper will focus on the other phases. As we will show, in these cases, the leading temperature dependence of $\lambda_L$ can be calculated by perturbing an effectively non-interacting low temperature limit, where the ratio between boson and fermion numbers vanishes. 
	This stands in contrast with the fast scramblers, whose maximal Lyapunov exponent is a consequence of a non-perturbative conformal solution~\cite{Kim:2019lwh}. 
	But while the calculation relies on a perturbative solution for the Green's functions, the behavior of the OTOC can be non-analytic in the coupling constant and reflect the universal low energy properties of the state (see Table~\ref{tab:summary}). For Fermi liquids, we find that $\lambda_L$ has the same $T$-dependence as the quasiparticle decay rate $1/\tau \ll T$ and it is also proportional to the small coupling constant. In the non-Fermi liquids we considered,  quasiparticles are over damped, having a decay rate $1/\tau \gg T$. Then we find $\lambda_L = \alpha 2\pi T $ is linear in $T$, and the prefactor $\alpha$ is independent of the coupling constant even as it approaches zero.  For the marginal Fermi liquid, we find that $\lambda_L$ is $T$-linear, yet the prefactor vanishes in a non-analytical fashion in the non-interacting limit [see 
	also \eqref{eq:marginal_result} below]:
	\begin{equation}
	\lambda_L \sim T g \ln (1/g) \,, \label{eq:marginal_intro}
	\end{equation}
	where $g$ is the coupling constant. The bulk of this paper will consist of a detailed derivation of the above results (Section~\ref{sec:lowrank}), preceded by a warm-up exercise where we analyze the mass-deformed SYK model, which is a simpler large-$N$ model of a disordered Fermi liquid.

	\section{Warm-up: mass deformed SYK}\label{sec:2+4}
	In this section, we revisit scrambling in the mass-deformed SYK model, obtained by adding all-to-all, quadratic interactions to the standard SYK model:
	\begin{equation}
	\begin{split}
	H &=  \sum_{i,j = 1}^N \im J_{2,ij}\gamma^i\gamma^j +    \sum_{i,j,k,l=1}^N J_{4,ijkl}\gamma^i\gamma^j\gamma^k\gamma^l \,. 
	\label{eq:SYK24Ham}
	\end{split}
	\end{equation} 
	Here $\gamma_1, \dots \gamma_N $ are Majorana fermions satisfying the anticommutation relations $\gamma_i \gamma_j + \gamma_j \gamma_i = \delta_{ij}$, and $J_{2,ij}, J_{4,ijkl}$ are independent random couplings, each with a centered Gaussian distribution of variance:
	\begin{equation}
	\braket{J_{2,ij}^2} = \kappa^2/N \,,\, \braket{J_{4,ijkl}^2} = 6J^2/N^3 \,.
	\end{equation}
	The result of the calculation, Eq.~\eqref{eq:lambdaL_24} below, has been reported by some of us in \cite{Kim:2020mho}. Our goal here is to illustrate the methods in a simpler setting. 
	In sections ~\ref{sec:basics} and \ref{sec:Lya} respectively, we review the basic approach to obtain the single particle Green's functions and the Lyapunov exponent by analyzing the ladder kernel. In Section~\ref{subsec:SYK24Perturbation}, we compute $\lambda_L$ using a perturbation theory around a non-interacting limit. Finally, we discuss the physical interpretation in Section~\ref{sec:2+4discussion}. 
	
	\subsection{Basics}\label{sec:basics}
	
	In the large-$N$ limit, the model (\ref{eq:SYK24Ham}) can be solved exactly by dynamical mean field theory: the disorder averaged fermion Green function $G$ and self energy $\Sigma$ satisfy a closed set of Schwinger-Dyson (SD) equations, whose real-time form is as follows:
	\begin{align}
	& \Sigma_{>}(t) = J^2 G_{>}(t)^3 \,,\, 
	\Sigma_{<}(t) = J^2 G_{<}(t)^3 \,, \label{eq:SD_Sigma_24} \\
	&G_R(\omega) = \frac1{\omega - \kappa^2 G_R(\omega) - \Sigma_R(\omega)} \,.  \label{eq:SD_G_24} 
	\end{align}
	Here, the various components of the Green function are defined as  
	\begin{align}
	&G_>(t) \equiv  \left< \gamma_1(t) \gamma_1(0) \right> \,,\, 
	G_<(t) \equiv  \left< \gamma_1(0) \gamma_1(t)  \right> \,,\,  \\
	&    G_R(t) \equiv -\im \theta(t) \left<  \gamma_1(t) \gamma_1(0) + \gamma_1(0) \gamma_1(t)  \right>  \,,
	\end{align}
	and $\left< [\dots] \right>$ denotes both thermal and disorder average. Throughout this paper, we use a convention in which
	\begin{itemize}
		\item $G_>(t) = G_E(\im t + 0)$, $G_<(t) = G_E(\im t + \beta - 0)$ are the analytical continuation of the time-ordered Euclidean Green function;
		\item  $G_R(\omega \to -\im \omega_n) = G_E(\omega_n)$ for Matsubara frequencies $\omega_n < 0$.
	\end{itemize}
	Therefore, real-time SD equations can be easily obtained by continuing the Euclidean ones. At thermal equilibrium of inverse temperature $\beta$, the fermion Green functions satisfy the fluctuation-dissipation relations:
	\begin{equation}
	G_{\lessgtr}(\omega) (1+e^{\pm\beta \omega} ) = -2 \Im \, G_R(\omega) \,. \label{eq:FDT}
	\end{equation}
	For computing the OTOC, we shall also make use of the Wightman correlator, defined as 
	\begin{equation}
	G_{lr}(t) \equiv \left< \gamma_1(\im\beta/2 + t) \gamma_1(0) \right>  \,. 
	\end{equation}

	\subsection{Ladder diagram and Bethe-Salpeter equation}\label{sec:Lya}
	We consider the Lyapunov exponent associated with the OTOC of the fermion operator, defined in the following standard manner:
	\begin{align}\label{eq:OTOC_reg}
	C(t) = \left< \{\gamma_2(t + \im \beta/2 ), \gamma_1(\im\beta/2)\}^\dagger \{\gamma_2(t), \gamma_1(0)\} \right> \,.
	\end{align}
	Compared to \eqref{eq:OTOC_intro}, the commutator is replaced by an anti-commutator as the observables are fermionic. Also, for regularization purposes, the operators are displaced along the thermal circle; it has been argued that this has no effect on the Lyapunov exponent in SYK-type models~\cite{kobrin2020many}. 
	
\begin{figure}
		\centering
		\includegraphics[width = 0.7\columnwidth]{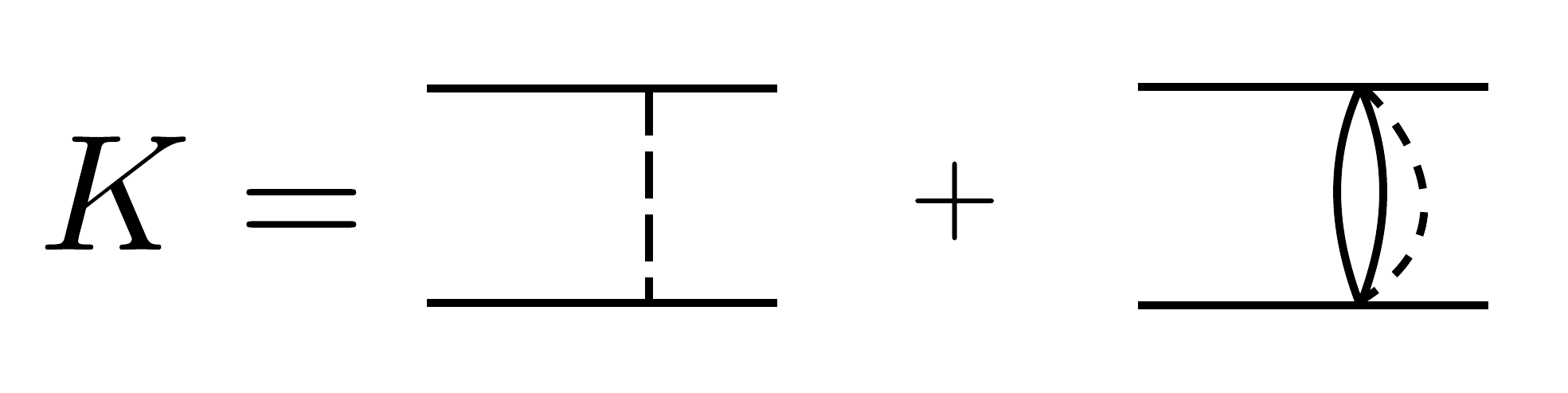}
		\caption{The ladder kernel for the mass-deformed SYK model: it is a sum of two kernels, as each rung in the ladder diagrams can be of two types. A solid line represents a fermion propagator, and a dashed line a disorder contraction. Note that there are no propagators associated with the short horizontal lines.}
		\label{fig:kernel_SYK24}
	\end{figure}

	In a $1/N$ expansion, the leading contribution to the OTOC is of order $1/N$, and comes from an infinite series of ladder diagrams. 
	There is a simple scheme, proposed by Kitaev~\cite{Kitaev:2015,syk}, to directly calculate $\lambda_L$ from the ladder diagrams. 
	The first step is to identify an integral kernel that generates the ladder diagrams. The action of the kernel adds a rung to the ladder as illustrated in Fig.~\ref{fig:kernel_SYK24} for the kernel of the mass-deformed SYK. In general, the kernel maps a function $F(t_1, t_2)$ of two time variables to another one: 
	\begin{equation}
	(KF)(t_3, t_4) = \int dt_1 dt_2 K(t_1, \dots, t_4) F(t_1,t_2) \,. \label{eq:Kaction}
	\end{equation} 
	The precise form $K(t_1, \dots, t_4)$ should be read off from the diagram; in particular, a  propagator in the longitudinal (transverse) direction should be $\im G_R$ ($G_{lr}$, respectively). Then, we restrict to the following space of factorized functions~\footnote{For any $\lambda_L$, this space is closed under the action of $K$, so long as $K(t_1, \dots, t_4) =K(t_1 + \delta t, \dots, t_4+\delta t) $ is invariant under time translation.}:
	\begin{equation}
	F(t_1, t_2) = e^{\lambda_L (t_1 + t_2)/2} f(t_1 - t_2) \,. \label{eq:growth_ansatz}
	\end{equation} 
	Here, $\lambda_L$ should be regarded as a free parameter to be determined, and $K$ a $\lambda_L$-dependent linear map that acts on functions of one variable $f(t) \mapsto (Kf)(t)$. Finally, we determine $\lambda_L$ by requiring that the most positive eigenvalue of $K$ be equal to $1$ (in general, $K$ is conjugate to a Hermitian operator, so has a real spectrum). Intuitively, this procedure searches for $\lambda_L$ self-consistently, by requiring that adding a rung to the ladder does not change it. Therefore, the scheme is effectively summing all ladder diagrams and is non-perturbative in nature. The integral equation that defines the eigenvalue problem for $K$ is reminiscent of the Bethe-Salpeter equation and often referred to as such. 
	
	In the case of the mass-deformed SYK model, the ladder kernel can be read off from the diagrams Fig.~\ref{fig:kernel_SYK24}:
	\begin{equation*}
	K(t_{1,\dots,4}) = -G_R(t_{31})  G_R(t_{42}) \left[ \kappa^2 +  3 J^2 G_{lr}(t_{34})^2 \right] \,,
	\end{equation*}
	where $t_{ij} \equiv t_i - t_j$. Its action on a function of type  \eqref{eq:growth_ansatz} can be represented as the product of two kernels:
	\begin{subequations} \label{eq:K}
		\begin{align}
		&   K = K_1 K_2 \,,\, \text{where} \\  
		&  (K_1 f)(t) = \left[ \kappa^2 + 3 J^2 G_{lr}^2 (t) \right]f(t) \,, \\
		&  (K_2 f)(\omega) = 
		\left|G_R\left(\omega + \im \frac{\lambda_L}{2}\right)\right|^2f(\omega) \,.
		\end{align}
	\end{subequations}
	That is, $K_1$, which involves Wightman correlators, is diagonal in the time basis, while $K_2$, which involves retarded ones, is diagonal in the frequency basis. 
	
	Therefore, calculating $\lambda_L$ in the mass-deformed SYK model requires the knowledge of $|G_R(\omega)|^2$ and $G_{lr}$, to which we turn now.
	
	\subsection{Perturbative analysis}\label{subsec:SYK24Perturbation}
	To calculate the Green functions at arbitrary temperature, one needs to solve the SD equations numerically. However, at low temperatures, the quadratic coupling is more relevant than the quartic one, so we can proceed with a perturbative expansion in $J/\kappa$ from the non-interacting limit. We will carry out the expansion up to first order $\BigO(J/\kappa)$ (and in the low energy limit), which is sufficient for our purposes.
	
	At zero-th order, $\Sigma = 0$, and we have the conformal solution of SYK$_{q=2}$, where the fermions have scaling dimension $\Delta = 1/2$: 
	\begin{align}
	& G_{lr} (t) = \frac{T}{\kappa} \, \sech( \pi t T ) \,, \label{eq:G_lr_24} \\
	& G_R (\omega) = -\frac{\im}{\kappa}  \text{  (zero-th order)} \,. \label{eq:GR_0_24}
	\end{align}
	Plugging the above into the SD equation \eqref{eq:SD_Sigma_24}, we obtain the self-energy at first order in $J/\kappa$,
	\begin{align}
	&  -  \Im \,  \Sigma_R(\omega) =\frac12 ( \Sigma_>(\omega) + \Sigma_<(\omega)) \nonumber \\ 
	&= \frac4{\kappa^3}\int \frac{d\omega_1d\omega_2}{(2\pi)^2} n(\omega_1) n(\omega_2) n(\omega-\omega_1-\omega_2) + [\omega \to -\omega]  \nonumber \\
	&=  \frac{J^2 T^2}{2 \kappa^3} +\frac{J^2 \omega^2}{2\pi^2 \kappa^3} \,. \label{eq:sigma}
	\end{align} 
	where we denoted $n(\omega) \equiv (1+e^{-\beta \omega})^{-1}$. We will not need $\Re \Sigma_R(\omega)$, because $\Im \Sigma_R(\omega)$ alone determines $|G_R(\omega)|$ at low frequencies. This follows from the SD equation~\eqref{eq:SD_G_24}, considered as a quadratic equation of $G_R(\omega)$ (note that we treat the SYK$_{q=2}$ term non-perturbatively): its roots lie on the unit circle when $\omega - \Sigma(\omega)$ is real (and small enough). Otherwise, $G_R(\omega)$ is the root inside the unit circle, whose norm is the following
	\begin{align}
	& \kappa^2  \left|G_R\left(\omega + \im s \right)\right|^2 = 1 -  \frac1\kappa \mathrm{Im} (\omega +\im s - \Sigma_R(\omega + \im s)) \nonumber \\
	=&  1-  \frac{s}{\kappa}-\frac{J^2 T^2}{2\kappa^4} - \frac{J^2}{2\pi^2 \kappa^4}\omega^2 \,, \label{eq:GR2_24}
	\end{align}
	to the leading order in $s, J$ and $T$.   
	
	Plugging the perturbative results \eqref{eq:G_lr_24} and \eqref{eq:GR2_24} into the ladder kernel~\eqref{eq:K}, we obtain its expression at first order perturbation. It will be useful to write that as a sum of various contributions:
	\begin{equation}
	K = K_{\text{conf}} + K_{\text{kin}} + K_{\text{decay}} + K_{\text{int}} \,, \label{eq:K-decomp}
	\end{equation}
	where each term is diagonal either in $t$ or in $\omega$, as follows:
	\begin{align}
	&K_{\text{conf}}f = f \,,\, \\ 
	&K_{\text{kin}} f = -\frac{\lambda_L}{2\kappa} f \,,\\
	& (K_{\text{decay}}f)(\omega) = \left(\frac{J^2 T^2}{2\kappa^4} - \frac{J^2}{2\pi^2 \kappa^4}\omega^2  \right)f(\omega) \,,\, \\
	& (K_{\text{int}}f)(t) = \frac{3 J^2 }{\kappa^2} G_{lr}(t)^2  f(t) \,.
	\end{align}
	We will discuss their physical meanings below, and proceed now to calculate $\lambda_L$. Since $\omega = -\im \partial_t$, we can write the entire kernel in the time domain:
	\begin{align}
	K 
	& = 1- \frac{\lambda_L}{2\kappa} - \frac{J^2 T^2}{2\kappa^4} -  \nonumber \\
	& \frac{J^2T^2}{\kappa^4}\left(- \frac{1}{2\pi^2 T^2} \partial_t^2 - 3  \sech( t \pi T)^{2}  \right)\,. 
	\end{align}
	Therefore, up to a shift and a rescaling, the Bethe-Salpeter equation is reduced to a time-independent Schr\"odinger equation with a P\"oschl–Teller potential. The most positive eigenvalue of $K$ corresponds to the ground state energy. This is an exactly solved problem: one may check that the eigenfunction and eigenvalue are respectively
	\begin{align}
	& f(t) \propto \sech(t \pi T)^2 \,,\, k = 1-  \frac{\lambda_L}{2\kappa} + \frac{3J^2 T^2}{2\kappa^4}  \,. 
	\end{align}
	Setting $k$ to $1$, we obtain
	\begin{equation}
	\lambda_L = \frac{3T^2 J^2}{\kappa^3} \,.\label{eq:lambdaL_24}
	\end{equation}
	This prediction is exact in the low-temperature ($T/\kappa \ll 1$) and weak-interaction ($J/\kappa \ll 1$) regime. Since the interaction is irrelevant, we expect it to be valid even if $J / \kappa$ is not small, provided the temperature is low enough. In \cite{Kim:2020mho}, we verified \eqref{eq:lambdaL_24} with non-perturbative numerical calculations, and found a good agreement in a wide parameter range. 
	
	\subsection{Discussion}\label{sec:2+4discussion}
	The above analysis provides the mathematical underpinnings of the temperature dependence $\lambda_L \sim T^2$ that has been observed in large-$N$ Fermi liquids~\cite{sumilan,guo}. Let us now discuss the different contributions to the kernel $K$:
	\begin{itemize}[leftmargin=.2in]
		\item $K_{\text{conf}} = 1$ is the non-interacting conformal contribution. Indeed, with $G_R(\omega) = -\im/\kappa$ and $J =0$, the kernel $K$ is equal to identity, and does not determine $\lambda_L$: for that, we must take into account the next-order correction.  This is in stark contrast with fast scramblers such as the SYK$_q$ model (with $q > 2$), where $\lambda_L = 2\pi T$ results from the conformal-limit kernel alone. 
		\item $K_{\text{kin}}= - \lambda_L / (2 \kappa)$ is the ``kinetic'' contribution, since it originates from the term $\omega = -\im \partial_t$ in the SD equation \eqref{eq:SD_G_24}. At the leading order, it is the only contribution where $\lambda_L$ appears. 
		\item The ``decay'' contribution $K_{\text{decay}}$
		originates from the quasiparticle decay rate $-\Im \Sigma_R(\omega)$.
		\item Finally, the ``interaction'' contribution $K_{\text{int}}$ involves the Wightman correlator connecting the interaction verices. 
	\end{itemize}
	It is worth noting that quasiparticle decay alone would have hindered scrambling, if not for the interaction, which promotes it. In terms of the Schr\"odinger equation analogy, $K_{\text{decay}}$ is the positive definite kinetic term plus a positive constant shift, and $K_{\text{int}}$ is a negative potential. The dominant eigenfunction of $K$ corresponds to the ground state, which is a normalizable bound state, and $\lambda_L$ is essentially its energy. However, due to the constant shift in $ K_{\text{decay}}$, it is not a priori clear that $\lambda_L > 0$ (or equivalently, the bound state energy is negative). To verify that, we were obliged to explicitly find its value. On the other hand, its temperature dependence could have been guessed by a power counting. 
	
	As we shall see in the following section, the above structures are essentially retained in the Fermi-liquid states realized by the low-rank SYK model, except that the quasiparticle decay rate will scale as $T^{1+\eta}$, with $1 < 1 + \eta < 2$, instead of $T^2$, which will entail that $\lambda_L \sim T^{1+\eta}$ as well. In contrast, crucial differences will occur in the non-Fermi-liquid states.

	\section{Low-rank SYK}\label{sec:lowrank}
	We now turn to analyzing scrambling in the low-rank SYK model, following the approach illustrated above: after setting the stage in Sections~\ref{sec:lowrank_basics} and \ref{sec:lowrank_ladder}, we perform a perturbative analysis of the Fermi liquid states (class I in~\cite{Kim:2019lwh}) in Section~\ref{sec:lowrank-pert-1} and non-Fermi liquid states (class II) in Section~\ref{sec:lowrank-pert-2}. Finally, we discuss the marginal Fermi liquid state that can be found on the border. 
	
	\subsection{Schwinger-Dyson equations}\label{sec:lowrank_basics}
	Let us recall from the Introduction that the low-rank SYK model can be defined as a quantum mechanical system with a large number of Majorana fermions $\gamma^1, \dots, \gamma^N$ and real bosons $\phi_1, \dots, \phi_R$, interacting via random Yukawa couplings~\footnote{We have omitted the boson kinetic term, which would be irrelevant in any case~\cite{Kim:2019lwh}.}:
	\begin{equation}
	H = \sum_{n=1}^{R} \left[ \frac12 \phi_n^2 + \sqrt{\lambda_n} \phi_n \sum_{i,j=1}^N \im u_{ij}^{(n)}\gamma^i\gamma^j  \right]      \,.
	\label{eq:lowrankH}
	\end{equation}
	Here $u_{ij}^{(n)}$ are independent real Gaussian random variable with zero mean and $1/N^2$ variance, and $\lambda_1, \dots, \lambda_R$ are $\BigO(1)$ coupling constants that tend to a smooth distribution
	\begin{equation}
	\rho(\lambda) \equiv \frac1N \sum_{n = 1}^R \delta(\lambda-\lambda_n)
	\end{equation}
	in the thermodynamic limit. A key feature of this model is that~\cite{Kim:2019lwh}, its low-energy behaviors depend on the the shape of $\rho(\lambda)$ (which is an input parameter), or more precisely, on the shape near the right edge $\lambda \sim \lmax$ when $\lmax > 0$~($\lmax \le 0$ leads to a fast scrambler~\cite{Bi:2017yvx,Kim:2019lwh} that we will not consider here). This is because the boson modes $\phi_n$ with $\lambda_n \sim \lmax$ have the lowest effective masses $m \propto \lmax - \lambda_n$, so the edge $\rho(\lambda \sim \lmax)$ determines the leading low-frequency behavior of the sum of boson propagators that control the fermion decay. In what follows, we shall focus on the distributions whose right edge is at $\lmax > 0$ and characterized by a power-law singularity:
	\begin{equation}
	\rho(\lambda) = \gamma \, (1 - \lambda)^\eta \theta(1-\lambda) \,,\, \lambda \nearrow \lmax\equiv 1 \,. \label{eq:rho}
	\end{equation}
	Here, $\eta$ is the exponent which crucially affects the low-energy behavior: $\eta \in (0,1)$ leads to Fermi liquids, and $\eta \in (-1,0)$ non-Fermi liquids. $\gamma$ is a dimensionless coefficient that controls the number of boson modes (per fermion) near $\lambda = \lmax$. It will serve as the coupling constant and a small parameter in the perturbative analysis. We also set $\lmax = 1$ to alleviate notations (it will be restored by dimensional analysis in the final results).
	
	The above model is solveable in the large-$N$ limit; the SD equations have been derived and solved in Euclidean time in \cite{Kim:2019lwh}. To recall and transcribe them into real time, we define the boson propagators as per the convention in Section~\ref{sec:basics}:
	\begin{align}
	&G_{\lambda_n}^>(t) = \braket{\phi_n(t)\phi_n(0)} \,,\, 
	G_{\lambda_n}^<(t) = \braket{\phi_n(0)\phi_n(t)} \\
	&G_{\lambda_n}^R(t) = \im \theta(t) \braket{[\phi_n(t), \phi_n(0)]} \,,
	\end{align}
	In the thermodynamic limit, the boson modes follow a continuous distribution, so we can speak about $G_\lambda$ as a function of $\lambda$, and transform sums over boson modes into integrals over $\rho(\lambda) d\lambda$. 
	For instance, it will be convenient to introduce the following weighted sum of boson propagators:
	\begin{equation}
	F(t) = \int G_{\lambda}(t) \rho(\lambda) \lambda d\lambda    
	\label{eq:F}
	\end{equation}
	Note however that $F$ does not contain the contribution of an eventual condensate. Indeed, the boson mode(s) with $\lambda = \lmax = 1$ can condense, and generate an SYK$_{q=2}$ interaction. We shall denote its coupling constant as $\kappa$, adopting the notation from Section~\ref{sec:2+4} ($\kappa^2 = 2 \Phi$ where $\Phi$ is the condensate fraction defined in \cite{Kim:2019lwh}). With the above considerations in mind, the real-time SD equations are as follows:
	\begin{subequations}
		\begin{align}
		& \Pi_{\lessgtr}(t) = G_{\lessgtr}(t)^2 \label{eq:lowrankSD_Pi} \,, \\
		& G^R_{\lambda}(\omega) = \frac{1}{1- \lambda \Pi_R(\omega)} \,,\,  \label{eq:lowrankSD_boson} \\
		& G_R(\omega) = \frac{1}{\omega- \kappa^2 G_R(\omega) - \Sigma_R(\omega)} \label{eq:lowrankSD_G} \\ 
		& \Sigma_\lessgtr(t) = 2 G_\lessgtr(t) F_\lessgtr(t)  \,,
		\label{eq:lowrankSD_Sigma}
		\end{align}
	\end{subequations}
	Above, we introduced $\Pi$ such that $\lambda \Pi$ is the self energy of bosons with $\lambda_n = \lambda$. Finally, when $\kappa > 0$, we have also the condensation condition:
	\begin{equation}
	\Pi_R(\omega = 0) = 1 \,. \label{eq:condense}
	\end{equation}
	
	\subsection{Ladder kernels}\label{sec:lowrank_ladder}
	\begin{figure}
		\centering
		\includegraphics[width = 0.75\columnwidth]{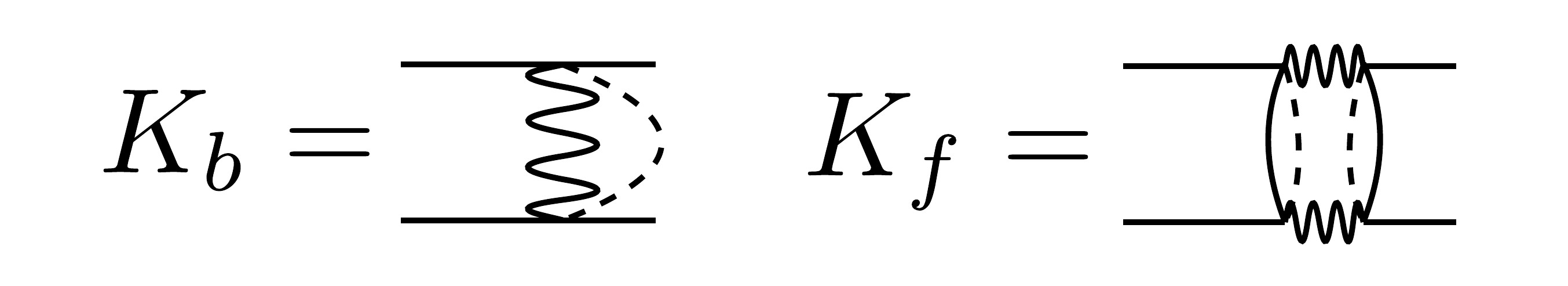}
		\caption{The two ladder kernels of the low-rank SYK model. A wavy line denotes a boson propagator. }
		\label{fig:ladders}
	\end{figure}
	We now turn to review the general form of the Bethe-Salpeter equations that we shall analyze to find $\lambda_L$. The ladder kernels of the low-rank SYK model is a sum of the bosonic one $K_b$ and the fermionic $K_f$, see Fig.~\ref{fig:ladders}~\cite{Patel1844,Kim:2019lwh}.  For each kernel, we can express its action on functions of type $F(t_1, t_2) = e^{\lambda_L t/2} f(t_1 - t_2)$ in terms of the product of a few factors, each of which is diagonal either in the time basis or in the frequency basis. The bosonic kernel is reminiscent of the ladder kernel of the mass-deformed SYK:
	\begin{subequations} \label{eq:K_b}
		\begin{align}
		& K_b = K_1 K_2 \,,\, \text{where} \\
		& (K_1 f)(t) = (\kappa^2 + 2 F_{lr}(t)) f(t) \\
		& (K_2 f)(\omega) = \left|G_R\left(\omega + \im \frac{\lambda_L}{2}\right)\right|^2 f(\omega) \,.
		\end{align}
	\end{subequations}
	The fermionic kernel is slightly more involved: 
	\begin{subequations} \label{eq:K_f}
		\begin{align}
		& K_f = 4 K_3 K_4 K_3 K_2 \,,\, \text{where}  \\ 
		& (K_3 f)(t) = G_{lr}(t) f(t) \,, \\
		& (K_4 f)(\omega) =  f(\omega) \underbrace{ \int \lambda^2 |G^R_{\lambda}(\omega+\im \lambda_L/2)|^2 \rho(\lambda)\mathrm{d} \lambda}_{k_4(\omega + \im \lambda_L / 2)} \label{eq:K_4} \,.
		\end{align}
	\end{subequations}
	Here, $K_4$ involves a sum over boson's retarded propagators, to which the condensate mode does not contribute.
	
	Therefore, to analyze the kernel, we need to compute $|G_R|^2$, $F_{lr}$, $G_{lr}$, and the integral defining $K_4$. As we showed in \cite{Kim:2019lwh}, this can be done by a perturbative analysis, with $\gamma$ in Eq.~\eqref{eq:rho} as a small parameter. While the calculations are formally similar for Fermi, non-Fermi and marginal Fermi liquids, the physical differences warrant treating them in turn. 
	
	Before moving on, we remark that the kernel $K_b + K_f$ determines not only the Lyapunov exponent of the OTOC between fermions, but also also between bosons
	\begin{equation*}
	\left< [\phi_n(t + \im \beta/2 ), \phi_m(\im\beta/2) ]^\dagger [\phi_n(t), \phi_m(0)] \right> \,,
	\end{equation*}
	or between a boson and a fermion. This is because the ladder diagrams of all these OTOCs are generated by the following generalized kernel:
	\begin{equation}
	\mathbf{K} \begin{pmatrix} f \\ 
	f_{\lambda} \end{pmatrix} 
	=  \begin{pmatrix}
	K_b f & + & \int  2 K_3  f_\lambda d \lambda  \\
	(2 K_{4, \lambda} K_3 K_2) f & + & 0  
	\end{pmatrix} 
	\end{equation}
	where 
	\begin{equation}
	(K_{4,\lambda} f)(\omega) =  f(\omega)   \lambda^2 |G^R_{\lambda}(\omega+\im \lambda_L/2)|^2 \rho(\lambda)  \,.
	\end{equation}
	The kernel $\mathbf{K}$ transforms between fermion and boson pairs propagating along the ladder (the boson pair appears in the middle of $K_f$, see Fig.~\ref{fig:ladders}), by acting on vector-valued functions $(f(t), \{f_\lambda (t)\}_{\lambda})$. The type of OTOC only affects the boundary of the ladders. 
	Now, it is straightforward to check that $\mathbf{K}$ and $(K_f + K_b)$ are related as follows:
	\begin{equation}
	\mathbf{K} \begin{pmatrix} f \\ 
	f_{\lambda} \end{pmatrix} =  \begin{pmatrix} f \\ 
	f_{\lambda} \end{pmatrix} \Leftrightarrow  (K_b + K_f) f = f \,,
	\end{equation}
	with $f_\lambda = 2 K_{4, \lambda} K_3 K_2 f$.  In other words, $1$ is an eigenvalue of $\mathbf{K}$ if and only if it is of $K_b + K_f$. As a consequence, by Kitaev's method, the Lyapunov exponent is the same for any OTOC, and can be computed by analyzing $K_b + K_f$.
	
	\subsection{Fermi liquids}\label{sec:lowrank-pert-1}
	We first analyze the Fermi liquid states (class I in \cite{Kim:2019lwh}). The perturbation scheme described above is controlled at low temperatures, because a boson condensate is always present ($\kappa > 0$), and the quadratic term generated is more relevant than the interaction mediated by the normal boson modes. In general, the value of $\kappa$ depends on $T, \gamma$ and $\eta$; for our analysis, it is unnecessary to know its precise value, other than the fact it has a nonzero finite limit $\kappa \to 4/(3\pi)$ ($\kappa \to 4\lmax/(3\pi)$ if we restore the units) as $\gamma \to 0, T\to 0$~\cite{Kim:2019lwh}. In fact, throughout our analysis, we will treat $\kappa$ as an independent variable. 
	
	\subsubsection{Zero-th order}
	At zero-th order, $\gamma = 0$ (this means that the ratio between boson and fermion number is vanishing in the thermodynamic limit, not that there are no bosons), the model reduces to SYK$_{q=2}$, so that \eqref{eq:G_lr_24} and \eqref{eq:GR_0_24} hold\textit{ verbatim}:
	\begin{align}
	&   G_{lr}(t) = \frac{T}{\kappa}\sech{( \widetilde{t}/2)} \,,\, 
	\widetilde{t} \equiv \frac{2 \pi t}{\beta} \,,
	\label{eq:lowrankGlr} \\
	&   G_R(\omega) = -\frac{\im}{\kappa} \text{ (order $0$)}  \,.
	\label{eq:lowrankGs}
	\end{align}
	As before, it will be sufficient to know $G_{lr}$ up to order $0$. It will also be useful to have the time-domain expressions:
	\begin{equation}
	\begin{split}
	&  G_+(t) \equiv G_>(t) + G_<(t) = \frac{2\delta(t)}{\kappa}  \text{ (order $0$)} \,, \\
	&  G_-(t) \equiv G_>(t) - G_<(t) = \frac{2\im}{\kappa\beta\sinh\frac{\pi t}{\beta}} \text{ (order $0$)}  \,,
	\label{eq:lowrankGferm}
	\end{split}
	\end{equation}
	which one can obtain using the fluctuation-dissipation relation \eqref{eq:FDT}. The boson propagators and the fermion self energy both vanish at zero-th order.
	
	\subsubsection{Boson propagators}
	We now proceed to the first order. First, plugging \eqref{eq:lowrankGs} into the SD equation \eqref{eq:lowrankSD_Pi} for $\Pi_\lambda$ gives us:
	\begin{align}
	\Im \Pi_R(\omega) &= \frac12( \Pi^>_\lambda(\omega) - 
	\Pi^<_\lambda(\omega)) \nonumber \\
	&= \frac12 \int \frac{\mathrm{d}\omega'}{2\pi} n(\omega') n(\omega-\omega') - [\omega \to -\omega] \nonumber  \\
	& = \frac{\omega}{\kappa^2 \pi}  \,.
	\label{eq:G^2_R}
	\end{align}
	On the other hand, the leading small frequency behavior of $\Re (\Pi_R (\omega))$ is fixed by the condensation condition~\eqref{eq:condense}. We have therefore
	\begin{equation}
	\Pi_R(\omega) = 1 + \frac{\im\omega}{\kappa^2\pi} \label{eq:PiR}
	\end{equation}
	at first order. Combining this with \eqref{eq:F} and the SD equation for $G_\lambda$ \eqref{eq:lowrankSD_boson}, we may find
	\begin{equation*}
	\begin{split}
	F_R(\omega) &= \int\frac{\lambda\rho(\lambda)}{1-\lambda \Pi_R(\omega)} d\lambda \\
	&= C - \frac{\gamma}{s_\eta} \left(1-\Pi^R(\omega)\right)^\eta \\ &=  C -  \frac{\gamma}{s_\eta} \left(-\frac{\im\omega}{\pi\kappa^2}\right)^\eta \,,
	\end{split}
	\end{equation*}
	where we denoted
	\begin{equation}
	s_\eta \equiv \frac{\sin\pi\eta} {\pi} \,,
	\end{equation}
	and performed the integral over $\rho$ assuming its edge shape \eqref{eq:rho},  using the following asymptotic formula 
	\begin{equation}\label{eq:integral}
	\int_0  \frac{ x^\eta \mathrm{d} x}{z - x} = C -  s_\eta^{-1} (-z)^\eta \,,\, z \nearrow 0 \,.
	\end{equation}
	Above, $C$ denotes a constant that depends on details of the integrand away from the edge. However, it has no dynamic effect: Indeed, a constant shift in $F_R$ corresponds to a delta peak $\delta(\tau)$ in the Euclidean Green function $F_E(\tau)$. Because the fermion Green function $G_E(\tau)$ is antisymmetric, $G_E(\tau) \delta(\tau) = 0$, and $C$ has no effect on the fermion self energy $\Sigma_E(\tau) \propto F_E(\tau) G_E(\tau)$. Therefore we may remove the constant for this purpose and write:
	\begin{equation}
	F_R(\omega) = -\frac\gamma{s_\eta}  \left(-\frac{\im\omega}{\pi\kappa^2}\right)^\eta \,. \label{eq:FR}
	\end{equation}
	Note that a similar calculation applies to $k_4(\omega)$ in Eq.~\eqref{eq:K_4} for any $\omega \in \C$:
	\begin{align}
	k_4(\omega) &= \int \frac{\lambda^2 \rho(\lambda) \mathrm{d}\lambda}{(1-\lambda \Pi_R )(1-\lambda \Pi_R^*)}\label{eQ:k4_gen}  \\
	& =  \int \left[ \frac{1}{1-\lambda \Pi_R} - \frac{1}{1-\lambda \Pi_R^*} \right] \frac{\lambda \rho(\lambda) \mathrm{d}\lambda}{\Pi_R - \Pi_R^* } \nonumber\\
	& = \frac{\gamma (2\pi T)^{\eta-1} }{s_\eta} \frac{\pi\kappa^2}{2 \im \Re\widetilde{\omega}} \left[\left(\frac{\im\widetilde{\omega}^*}{\pi\kappa^2}\right)^\eta-\left(\frac{-\im\widetilde\omega}{\pi\kappa^2}\right)^\eta  \right]  \nonumber
	\end{align}
	where in the last step, we introduced the rescaled frequency 
	\begin{equation}
	\widetilde\omega \equiv \frac{ \omega}{2\pi T} \,.
	\end{equation}
	Eq.~\eqref{eQ:k4_gen} will be useful later as the most nontrivial building piece of the fermionic kernel.
	
	We now come back to use  \eqref{eq:FR} to compute the Wightman correlator $F_{lr}(t)$. For this, we analytically continue the Euclidean time Green function $F_E(\tau)$. Indeed, recalling that $F_E(-\omega_n) = F_R(\im \omega_n)$ for all Matsubara freqencies $-\omega_n = -2\pi n T < 0$, we have
	\begin{align}
	&F_E(\tau)  =  T \sum_{n=1}^{\infty} F_R(\im\omega_n)  2 \cos( \omega_n \tau)      \label{eq:lowrank_FE} \\
	=& \frac{2^{\eta} \pi \gamma T^{\eta+1}}{\sin\pi\eta \ \kappa^{2\eta}} \left[ \textrm{Li}_{-\eta}(e^{-\frac{2i \pi \tau}{\beta}}) + \textrm{Li}_{-\eta}(e^{\frac{2i\pi \tau}{\beta}})\right] \,,
	\nonumber
	\end{align}
	where $ \textrm{Li}_s(z) = \sum_{k=1}^\infty z^k k^{-s}$ is the polylogarithm. Continuing to $\tau = \im t + \beta/2$ then gives the Wightman correlator:
	\begin{equation}
	\begin{split}
	&     F_{lr}(t) = \,
	\frac{\gamma T^{\eta+1}}{s_\eta\kappa^{2\eta}}  V_\eta(\widetilde{t}) \,,\, \label{eq:lowrank_Flr}\\
	&      V_{\eta}(\widetilde{t}) \equiv 2^{\eta} \left[ -\textrm{Li}_{-\eta}\left(-e^{-\widetilde t}\right) - \textrm{Li}_{-\eta}\left(-e^{\widetilde t}\right) \right]  \,, 
	\end{split}
	\end{equation}
	where we recall that the rescaled time is defined as
	\begin{equation}
	\widetilde t \equiv \frac{ 2\pi t}{\beta} \,.
	\end{equation}
	For any $\eta \in (0,1)$, $F_{lr}(t)$ is positive and goes to zero as ${t} \to \pm\infty$. Note that the integrated boson propagator $F$ is \textit{not} conformal (otherwise, $V_\eta$ would be a hyperbolic secant to some power); this is in contrast with the fast scramblers realized by low-rank SYK~\cite{Bi:2017yvx,esterlis,Kim:2019lwh}.
	
	\subsubsection{Fermion self-energy}
	We now come to the fermion self energy $\Sigma_R$. The SD equation for the retarded self energy obtained directly from equation~\eqref{eq:lowrankSD_Sigma}  is:
	\begin{equation}
	\Sigma_R(t) = \im \theta(t) ( G_+ (t) F_+ (t) +  G_- (t) F_- (t) ) \label{eq:sigma_Rt}
	\end{equation}
	where $F_\pm(t) \equiv F_>(t) \pm F_<(t)$, and $G_\pm(t)$ are given in \eqref{eq:lowrankGferm}. Since $G_+(t) \propto \delta(t)$, it suffices to calculate $ F_+(t=0)$, which is equal to $2 F_E(\tau = 0)$ by definition:
	\begin{align}
	F_+(0) &=  2 F_E(0) = 4\sum_{n=1}^\infty F_R(\im \omega_n) \nonumber  \\
	&=   \frac{\gamma T^{\eta+1}}{s_\eta \kappa^{2\eta}} 2^{\eta+2} \zeta(-\eta) \,, \label{eq:lowrankF_+}
	\end{align}  
	where the sum is divergent and regularized by the Riemann zeta function. Meanwhile, $F_-(t)$ is the Fourier transform of $2 \Im F_R(\omega)$:
	\begin{equation}
	\begin{split}
	F_-(t) & = \int 2 \Im F_R(\omega) e^{-i\omega t} \frac{d\omega}{2\pi} 
	\\ &= -\im\frac{\gamma \Gamma(\eta+1)}{(\pi \kappa)^\eta}\frac{\sgn(t)}{|t|^{\eta+1}}\,.
	\label{eq:lowrankF_-}
	\end{split}
	\end{equation}
	Plugging \eqref{eq:lowrankGferm}, \eqref{eq:lowrankF_-}, and \eqref{eq:lowrankF_+} into \eqref{eq:sigma_Rt}, and performing the Fourier transform, we obtain 
	\begin{equation}
	- \textrm{Im} \Sigma_R(\omega)  =    \frac{\gamma T^{\eta+1}}{\kappa^{2\eta+2} s_\eta}  \sigma_\eta\left({\omega\over 2\pi T}\right) \,, \label{eq:SigmaR_lowrank}
	\end{equation}
	where the scaling function is
	\begin{equation}
	\sigma_\eta(x) = 2^{\eta+2}\textrm{Re}\left[\zeta\left(-\eta, \frac{1}{2} + \im x \right)\right] -2^{\eta+2}\zeta(-\eta) \,, \label{eq:scaling_GR}
	\end{equation}
	$\zeta(s, a)$ being the generalized Riemann zeta function. Consequently, the equivalent of \eqref{eq:GR2_24} in the low rank SYK model is
	\begin{equation}
	\begin{split}
	& \kappa^2 |G_R(\omega+\im s)|^2 = 1 - \frac{s}{\kappa} + \frac{\textrm{Im}\{\Sigma_R(\omega + \im s)\}}{\kappa} \\
	& \ \ = 1 - \frac{s}{\kappa} - \frac{ \gamma T^{\eta+1}}{s_\eta\kappa^{2+2\eta}} 
	\sigma_\eta\left({\omega+\im s \over 2\pi T}\right) \,.
	\label{eq:lowrankG2}
	\end{split}
	\end{equation}
	
	Note that, the intermediate objects $F_{\pm}$ suffer from UV divergences, which we zeta-regularized.  The end result is finite, and we verified it numerically.
	
	\subsubsection{Computing the Lyapunov exponent}
	We are now in a position to analyze the ladder kernels \eqref{eq:K_b} and \eqref{eq:K_f}, whose perturbative form is obtained by plugging in  \eqref{eq:lowrankGlr}, \eqref{eq:lowrank_Flr}, \eqref{eq:lowrankG2}, \eqref{eQ:k4_gen}, and expand up to first order in $\gamma$. The resulting total kernel can be decomposed similarly as done in \eqref{eq:K-decomp} for the mass-deformed SYK:
	\begin{equation}
	K_b+K_f = K_{\text{conf}} + K_{\text{kin}} + \frac{\gamma T^{\eta+1}}{\kappa^{2\eta+2}} \widetilde{K} \,. \label{eq:scaling_classI}
	\end{equation}
	where $K_{\text{conf}} = 1$, $K_{\text{kin}} = -\lambda_L / 2\kappa$, and 
	\begin{equation}
	\widetilde{K} =  \widetilde{K}_{2} + 
	\widetilde{K}_{1} + 4 \widetilde{K}_{3} \widetilde{K}_{4}\widetilde{K}_{3}   \,.
	\end{equation}
	Here, $\widetilde{K}_{i}$ is the rescaled version of $K_{i}$ (with all powers of $T,\kappa,$ and $\gamma$ factored out) in Section~\ref{sec:lowrank_ladder}, with the conformal and kinetic contributions subtracted:
	\begin{align}
	&( \widetilde{K}_{2} f)(\omega) =  -  \frac1{s_\eta} \sigma_\eta (\widetilde{\omega} + \im \widetilde{\lambda}_L / 2) f(\omega) \label{eq:K2_tilde}\\ 
	& (\widetilde{K}_{1} f)( t) = \frac{2}{s_\eta} V (\widetilde{t}) f({t})\\
	& (\widetilde{K}_{3}f)(t) = \sech(\widetilde{t} / 2) f({t}) \\
	& (\widetilde{K}_{4}f)({\omega}) =  \widetilde{k}_4(\widetilde{\omega} + \im \widetilde\lambda_L / 2) f({\omega})\end{align}
	where 
	\begin{align}
	\widetilde{k}_4(\widetilde{\omega}) = \frac{2^{\eta-1} \pi}{s_\eta} \frac{\left({\im\widetilde{\omega}^*}\right)^\eta-\left({-\im\widetilde\omega}\right)^\eta  }{2 \im \Re\widetilde{\omega}}  \,. \label{eq:k4_tilde}
	\end{align}
	Let us also recall $s_\eta \equiv \sin (\pi \eta) / \pi$, and the rescaled variables are defined as
	$$ \widetilde{t} = 2\pi T t \,,\,  \widetilde{\omega} = \frac{\omega}{2\pi T} \,,\, \widetilde\lambda_L = \frac{\lambda_L}{2\pi T} \,. $$  
	To further compare with the mass-deformed SYK case, $\widetilde{K}_2$ is the decay contribution which suppresses scrambling, while both $\widetilde{K}_1$ from the bosonic kernel and $4 \widetilde{K}_{3} \widetilde{K}_{4}\widetilde{K}_{3}$ (the fermionic kernel) promote it.

	Now, $\lambda_L$ is determined by $K_f + K_b$ having $1$ as the most positive eigenvalue. By  \eqref{eq:scaling_classI}, this amounts to
	\begin{equation}\label{eq:lambda_L_FL}
	\lambda_L = \frac{2 T^{\eta+1} \gamma}{\kappa^{2\eta+1}} \widetilde{k}(\widetilde{\lambda}_L) \,,
	\end{equation}
	where $\widetilde{k}$ is the most positive eigenvalue of $\widetilde{K}$, which depends a priori on $\widetilde{\lambda}_L$. However, Eq. \eqref{eq:lambda_L_FL} implies $\widetilde{\lambda}_L \propto T^{\eta}$ which vanishes as $T\to 0$. To leading order in the temperature dependence we can take $\widetilde{k}(0)$ on the right hand side as long as $\widetilde{k}(0) > 0$, which we shall check below. Thus, to leading order $\lambda_L$ is given by:
	\begin{equation}
	\lambda_L = \frac{2 T^{\eta+1} \gamma}{\kappa^{2\eta+1}} \widetilde{k}(0) \,.  \label{eq:LyapunovClassI_0}
	\end{equation}
	It remains to verify that $\widetilde{k}(0) > 0$: this is not a priori clear since $\widetilde{K}$ is the sum of a negative definite operator ($\widetilde{K}_2$, which originates from quasiparticle decay) and positive definite ones.   While we are not able to find $\widetilde{k}(0)$ analytically, it can be computed numerically to a high precision for any given $\eta \in (0,1)$. As a result, we found that the prefactor does not vanish for any $\eta \in (0,1)$. {We remark that to ensure $\widetilde{k}(0) > 0$, it is important to include the fermionic kernel contribution as well as the bosonic one. Indeed, we found numerically that the contribution of the bosonic kernel alone can be negative. }
	
\begin{figure}
		\centering
		\includegraphics[width = 1.0\columnwidth]{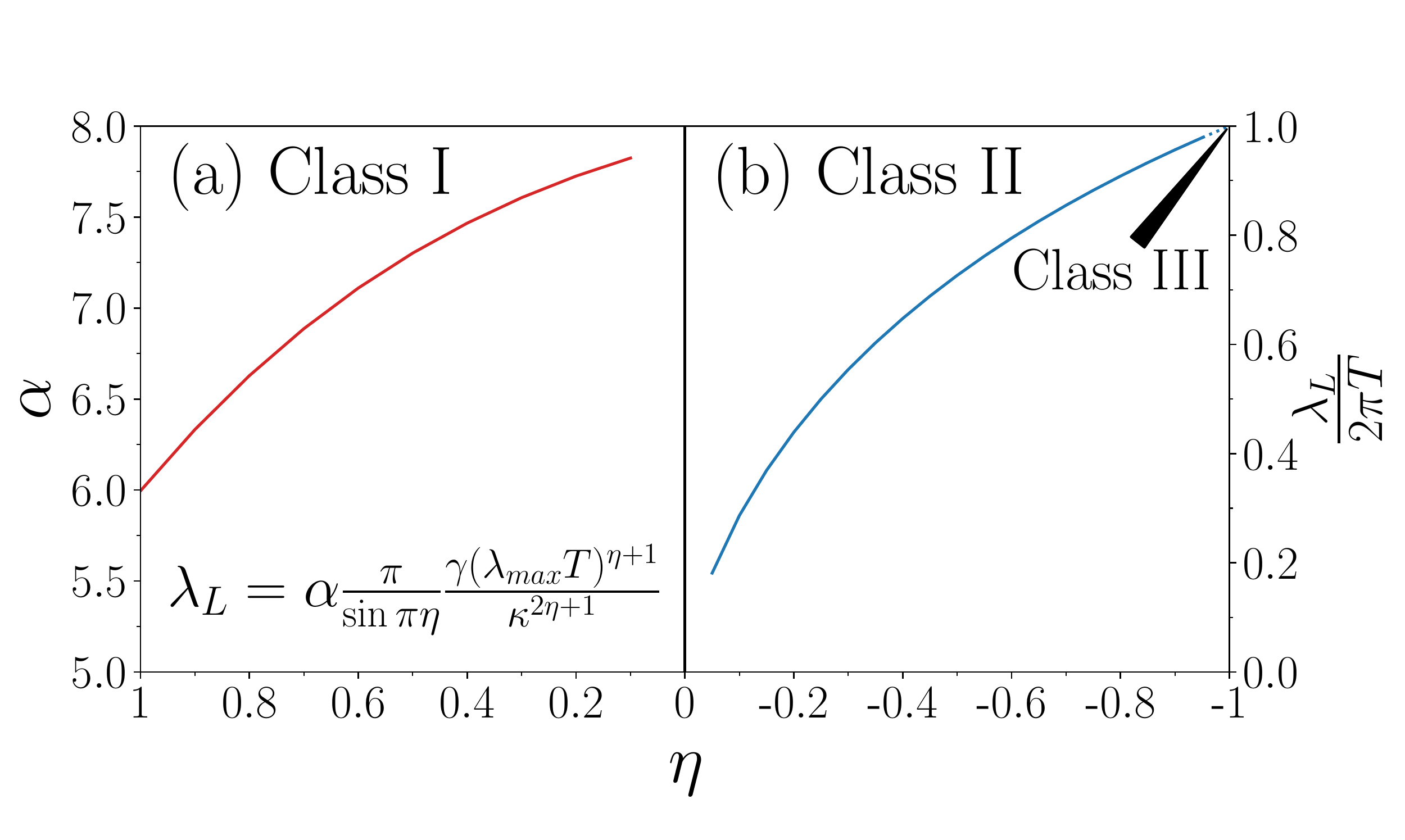}
		\caption{(a) The prefactor in the perturbative formula \eqref{eq:LyapunovClassI} for the Lyapunov exponent for the Fermi liquid states of low-rank SYK, as a function of $1>\eta>0 $. (b) the rescaled Lyapunov exponent ${\lambda_L}/({2\pi T})$ in the non-Fermi liquid states as a function of $0>\eta>-1$. Note that $\eta$ decreases along the $x$-axis. }
		\label{fig:lowrank_Lyapunov}
	\end{figure}

	Gathering all prefactors and restoring the dimensions, the Lyapunov exponent for the Fermi liquids realized in low-rank SYK is
	\begin{equation}
	\lambda_L = \alpha \frac{\pi}{\sin\pi\eta} \frac{\gamma (\lmax T)^{\eta+1}}{\kappa^{2\eta+1}}
	\label{eq:LyapunovClassI}
	\end{equation}
	where the numerically determined prefactor $\alpha$ is plotted in Fig.~\ref{fig:lowrank_Lyapunov} (a). We observe that $\alpha$ does not vary significantly as a function of $\eta$. More important is the factor $\sin \pi\eta$ that diverges in the limits $\eta \to 1$ and $\eta \to 0$. This is due to the fact that, for  $\eta = 0, 1$, the integral over $\rho(\lambda) = (1-\lambda)^\eta$ \eqref{eq:integral} will have additional log corrections. In particular, when $\eta = 0$, this log correction leads to a marginal Fermi liquid, which we deal with separately in section~\ref{sec:marginal} below. 
	
	\subsection{Non-Fermi liquids}\label{sec:lowrank-pert-2}
	We now turn to the non-Fermi liquid states (class II in \cite{Kim:2019lwh}). As we have demonstrated in \cite{Kim:2019lwh}, these states can be still accessed by a perturbative analysis that bears much formal resemblance to that of the Fermi liquid states . Therefore, we will be able to repeat most of the computations of the previous section, by properly reinterpreting them.  
	
	A distinguishing feature of the non-Fermi liquids~\cite{Kim:2019lwh} is that, there is no genuine condensate at finite temperature, so $\kappa = 0$. However, {at low temperatures}, the zero-frequency component of the soft boson modes generates an SYK$_{q=2}$ term which is again the most relevant. Its coupling constant will be denoted $\widehat{\kappa}$,  since it behaves similarly to the parameter $\kappa$ in the Fermi liquids; in particular, they have the same $\gamma \to 0$ limit. 
	
	Therefore, at zero-th order in $\gamma$, the fermion Green functions are still given by \eqref{eq:lowrankGlr}, \eqref{eq:lowrankGs}, \eqref{eq:lowrankGferm}, and $\Im \Pi_R(\omega)$ still satisfies \eqref{eq:G^2_R}, provided $\kappa$ is replaced by $\widehat{\kappa}$. 
	However, \eqref{eq:PiR} no longer holds; we have rather 
	\begin{equation}
	\Pi_R(\omega) = 1-\epsilon + \frac{\im \omega}{\widehat\kappa^2 \pi}
	\end{equation}
	with $\epsilon > 0$ so that 
	\begin{equation}
	F_R(\omega) = - \gamma \frac{\pi}{\sin\pi\eta} \left(\frac{\epsilon - \im\omega  }{\pi\widehat\kappa^2}\right)^\eta
	\end{equation}
	has a regularized singularity at $\omega = 0$ (recall that $\eta < 0$). In fact, the zero-frequency component of $F$ is responsible for the SYK$_{q=2}$ interaction $\propto \widehat\kappa^2$~(this was referred to as the effective condensate in \cite{Kim:2019lwh}):
	\begin{equation}
	2 T F_R(\omega = 0) = 2 T F_E(\omega_n = 0) = \widehat\kappa^2 \,, \label{eq:zero_freq}
	\end{equation}
	which implies that
	\begin{equation}
	\epsilon = c (T\gamma)^{-\frac1\eta} 
	\end{equation}
	is much smaller than $T$ at low temperatures  ($c$ is an unimportant constant depending on $\widehat{\kappa}$ and $\eta$).
	
	We can now take the analytic continuation to obtain the bosonic Whitman propagator
	\begin{equation}
	F_{lr} (t) = \frac{\widehat{\kappa}^2}{2} + \underbrace{\gamma T^{\eta+1}  s_\eta^{-1} V_\eta(\widetilde{t})}_{\widehat{F}_{lr}(t)}  \,. \label{eq:F_lr_classII}
	\end{equation}
	Note that it is of the same form as ~\eqref{eq:lowrank_Flr}, except for the  contribution of the zero Matsubara frequency. The latter parallels the condensate contribution to the nernel in \eqref{eq:K_b}(b).
	
	Similarly, the calculation of the fermion self-energy is the same as in the previous section except for the zero Matsubara frequency, which is now included exlicitly. Again, the latter mirrors the condensate contribution to the fermion Green's function with $\kappa$ now replaced by $\widehat{\kappa}$. Therefore, we have
	\begin{equation}
	\Sigma_R(\omega) = \widehat{\kappa}^2 G_R(\omega) + \widehat{\Sigma}_R(\omega) \,,
	\end{equation}
	where $\widehat{\Sigma}_R(\omega)$ has the same imaginary part as \eqref{eq:SigmaR_lowrank}:
	\begin{equation}
	- \Im \widehat{\Sigma}_R(\omega) =  \frac{ \gamma T^{\eta+1}}{s_\eta   \widehat\kappa } \sigma_\eta\left( \frac{\omega}{2\pi T} \right)  \,,
	\end{equation}
	with $\sigma_\eta$ is defined in \eqref{eq:scaling_GR}. 
	
	The effect of $\epsilon$ on $k_4(\omega + \im \lambda_L/2)$ is a shift of the argument $  \lambda_L  \to \lambda_L  + 2 \epsilon$, or $\widetilde\lambda_L \to \widetilde\lambda_L + \epsilon / (\pi T)$, which is negligible, except for $\widetilde\lambda_L = \BigO(T^{-1/\eta-1})$; we shall see that $\widetilde\lambda_L$ is much larger than that.
	
	Putting everything together we obtain the ladder kernel in the same form as in the Fermi liquids, only with $\kappa$ replaced by   $\widehat{\kappa}$:
	\begin{equation}
	K_b+K_f = K_{\text{conf}} + K_{\text{kin}} + \frac{\gamma T^{\eta+1}}{\widehat\kappa^{2\eta+1}} \widetilde{K} \,. \label{eq:scaling_classII}
	\end{equation}
	Here again $ K_{\text{conf}} = 1$, $K_{\text{kin}}  = -\lambda_L / (2\widehat{\kappa})$.
	The crucial difference from the Fermi liquids, however, is that because $\eta<0$ the last term in \eqref{eq:scaling_classII} is now dominant over 
	$K_{\text{kin}}$ due to the general bound $\lambda_L \le 2 \pi T$. Therefore, to leading order in $T$, we can drop $ K_{\text{kin}}$, so $\lambda_L$ is determined by requiring the most positive eigenvalue of $\widetilde{K}$ to be zero. 
	Since $\widetilde{K}$ only depends on $\eta$ and the rescaled exponent $\widetilde{\lambda}_L = \lambda_L /( 2\pi T) $, the Lyapunov exponent must be  $T$-linear, 
	\begin{equation}
	\lambda_L = a(\eta)\, T \,,
	\end{equation}
	with a prefactor that only depends on $\eta$. We computed the prefactor numerically by calculating the dominant eigenvalue of $\widetilde{K}(\widetilde{\lambda}_L)$ and applying a bisection search.
	The result is plotted in Fig.~\ref{fig:lowrank_Lyapunov} (b). We observe that $\widetilde\lambda_L$ is a continuously decreasing function of $\eta$. As $\eta \rightarrow -1$, $\widetilde{\lambda}_L \to 1$ comes arbitrary close to the universal bound, which is expected as the model approaches some fast-scrambling states (class III) of low-rank SYK \cite{Kim:2019lwh}. As $\eta \to 0$, $\widetilde{\lambda}_L$ stays positive while approaching zero. This is again related to the fact that the case $\eta = 0$ is a marginal Fermi liquid, which we turn to now.
	
	\subsection{Marginal Fermi liquid}\label{sec:marginal}
	We now consider the marginal Fermi liquid state in low-rank SYK, which is realized with $\eta = 0$. This means that the distribution $\rho (\lambda) = \gamma \theta(1-\lambda)$ is a step function near the edge. We cannot naively take a limit $\eta \to 0$ from either side, because the integral over $\rho(\lambda)$ [see \eqref{eq:integral}] acquires an additional log correction at $\eta = 0$. A scaling analysis in Euclidean time and at zero temperature yields $F_E(\omega) \sim -\ln |\omega|$, $F_E(\tau) \sim 1/|\tau|$, $\Sigma_E(\tau) \sim G_E(\tau) F_E(\tau) \sim 1/|\tau|^2 \mathrm{sign}(\tau)$ and thus 
	\begin{equation}\label{eq:SigmaE_marginal}
	\Sigma_E(\omega) \sim -\im \omega \ln |\omega| \,. 
	\end{equation} 
	The log correction means that we have a marginal Fermi liquid. At finite temperature, the system resembles the non-Fermi liquids in that there is no condensate but rather an effective condensate~\cite{Kim:2019lwh}.
	
	In spite of the apparent similarity to the non Fermi liquid case, computation of the ladder kernel turns out to be somewhat more involved for the marginal Fermi liquid. 
	Naively, one might guess that the non-trivial part of $K$ would have the same $T$ dependence as the quasiparticle decay rate, in which case:
	$$ K_b + K_f = K_{\text{conf}} + K_{\text{kin}} + \BigO\left( \gamma T \ln T \right) \,. $$ If this were the case, then the last term  would be dominant over $K_{\text{kin}}$ and the analysis would proceed as in the non-Fermi liquid. However, a more careful analysis shows that this is not quite true. Rather, for any $\widetilde\lambda_L > 0$, the nontrivial part of the kernel is just $T$-linear, without log correction, but has a singular dependence on $\widetilde\lambda_L$ when it goes to $0$. Taking this into account leads to the final result \eqref{eq:marginal_result}, which shows $\lambda_L\propto T=$, but with a prefactor that depends on the coupling constant $\gamma$ in a non-analytic way. This is different from either the Fermi liquid or the non-Fermi liquid results. 
	
	We now present the derivation of the Lyapunov exponent in  detail. To start, let us reconsider the integral \eqref{eq:integral}
	\begin{align}
	&I_\eta(z) \equiv \int_0  \frac{ x^\eta \mathrm{d} x}{z - x} \stackrel{z\sim 0}= s_\eta^{-1} (-z)^\eta   \,,\, \eta \ne 0  \,,
	\end{align}
	that enters the calculation of the Bosonic propagator $F_R(\omega)$. We recall that $ s_\eta \equiv {\sin\pi\eta}/{\pi}$ and note that for $\eta = 0$ the integral can be expressed as a limit of a derivative
	\begin{equation}
	I_0(z) = -\ln(-z) = \left. \partial_\eta \left[ s_\eta I_\eta(z)\right]\right\vert_{\eta\to 0}  \label{eq:eta-rel}\,.
	\end{equation}
	We will use this property to compute the nontrivial part of the ladder kernel 
	\begin{equation}
	K_\eta   \equiv  K_b + K_f - K_{\text{conf}} - K_{\text{kin}} \,,
	\end{equation}  
	which for $\eta<0$ is equal to 
	$$ K_\eta = \frac{\gamma T^{\eta+1}}{\widehat{\kappa}^{2\eta+2}} \widetilde{K} = 
	\frac{\gamma T^{\eta+1}}{\widehat{\kappa}^{2\eta+2}} \left( \widetilde{K}_1 + \widetilde{K}_2 + 4 \widetilde{K}_3 \widetilde{K}_4 \widetilde{K}_3\right)\,.$$  
	Crucially, because $K_\eta$ is a linear functional of $I_\eta(z)$ it satisfies the same relation~\eqref{eq:eta-rel}:
	\begin{align}
	K_0 = \partial_{\eta} \left[ \frac{s_\eta\gamma T^{\eta+1}}{\kappa^{2\eta+2}} \widetilde{K} \right]_{\eta \to 0}  \,.
	\end{align}
	The apparent leading $T$-dependence of $K_0$ comes from applying the partial derivative to the factor $T^{\eta+1}$ above; it is given by
	\begin{align}
	K_0 &= \frac{\gamma T\ln T}{\widehat\kappa^{2}}  \lim_{\eta\nearrow 0}[s_\eta \widetilde{K}] + \BigO(\gamma T) \, , \label{eq:K0-leading}
	\end{align}
	which was our naive expectation. However, it turns out  that the factor $\lim_{\eta\nearrow 0}[s_\eta \widetilde{K}]$ vanishes for any $\widetilde{\lambda}_L>0$. Let us show how this happens by inspecting the building blocks of $\widetilde{K}$ in the above limit. First
	For the fermionic kernel, \eqref{eq:k4_tilde} implies that 
	\begin{equation}
	s_\eta  \widetilde{k}_4(\widetilde{\omega} + \im \widetilde{\lambda}_L) \to \begin{dcases} 0   & \widetilde\lambda_L > 0 \\
	\frac\pi2 \delta(\widetilde{\omega})    & \widetilde\lambda_L = 0 \,.
	\end{dcases}
	\end{equation}
	Second, considering the bosonic Wightman correlator, we observe that $V_{\eta\to0} = 1$ from the defition~\eqref{eq:lowrank_Flr}. Hence in this limit we have
	\begin{equation}\label{eq:seta_K1}
	s_\eta \widetilde{K}_1  \to 2\, ,   
	\end{equation}
	proportional to an identity operator.
	Third, we consider the contribution $\widetilde{K}_2$ stemming from the decay rate $\text{Im}\Sigma_R$. 
	Inspecting the two contributions to the self energy in \eqref{eq:sigma_Rt} we observe that $s_\eta F_- \to 0$ in the limit $\eta\to 0$. This is because $F_-$ is an integral over $\Im [s_\eta F_R(\omega)] \to 0$, which vanishes as $\eta\to 0$ according to 
	\eqref{eq:FR}. Thus the only contribution to the decay rate comes from $F_+ G_+$. Thus we have
	\begin{equation}
	s_\eta \Im\Sigma_R(\omega) \to 4 \frac{\gamma T}{\widehat{\kappa}^2} (-\zeta(0)) = 2 \frac{\gamma T}{\widehat{\kappa}^2} \,
	\end{equation}
	and therefore
	\begin{equation}\label{eq:seta_K2}
	s_\eta \widetilde{K}_2 \to -2 \,.
	\end{equation}
	This cancels \eqref{eq:seta_K1} exactly. 
	
	The upshot of this analysis is that the leading ($T\ln T$) temperature dependence of the ladder kernel has a vanishing prefactor for  $\lambda_L>0$.
	We remark that this is consistent with the numerical observation in Fig.~\ref{fig:lowrank_Lyapunov} (b) that $\lambda_L/(2\pi T) \to 0$ as $\eta \to 0$ from the non-Fermi liquid side.  Of course, this does not mean that $\lambda_L$ vanishes in the marginal case, but rather that 
	to determine ${K}_0$ it is necessary to include the next to leading order in $T$, namely the term linear in $T$:
	\begin{equation}
	K_0 = \frac{\gamma T }{\widehat\kappa^{2}} \widetilde{K}_{\eta \to 0} \,,\, \widetilde{K}_{\eta \to 0}  \equiv  \partial_{\eta}\left[s_\eta  \widetilde{K} \right]_{\eta \to 0}\,,\,  \widetilde{\lambda}_L > 0 \,.
	\end{equation}
	
	We find that both $K_0$ and the kinetic contribution $K_{\text{kin}} =- \pi T\widetilde\lambda_L / \widehat{\kappa}$ are linear in $T$. So, unlike in the non-Fermi liquid case, the latter cannot be ignored. $\lambda_L$ is determined by requiring that $K_{\text{kin}}  + K_0$ have vanishing dominant eigenvalue, or, equivalently:
	\begin{equation} \label{eq:lambda_L:margimal}
	\lambda_L = \frac{2 T\gamma}{\widehat\kappa} \widetilde{k}_{\eta \to 0}(\widetilde\lambda_L) \,,
	\end{equation}
	where $\widetilde{k}_{\eta \to 0} $ denotes the dominant eigenvalue of $ \widetilde{K}_{\eta \to 0}$. Eq.~\eqref{eq:lambda_L:margimal} is reminiscent of \eqref{eq:lambda_L_FL} in the Fermi liquid case. In particular, it implies that $\lambda_L$ is linear in $T$. To obtain the prefactor to leading order in the coupling constant, one would be tempted to take $\widetilde\lambda_L\to 0$ on the RHS of \eqref{eq:lambda_L:margimal}, as done in the Fermi liquid case \eqref{eq:LyapunovClassI_0}. However, this approach fails because $\widetilde{k}_{\eta \to 0} (0) $ is divergent, with the culprit being the fermionic kernel, which has the following behavior
	\begin{equation}
	\begin{split}
	&  \partial_\eta \left[ s_\eta \widetilde{k}_4(\widetilde\omega + \im \widetilde\lambda_L/2) \right]_{\eta \to 0}
	\\ =& \frac\pi2 \frac{\Im\ln(\im\widetilde\omega +  \widetilde\lambda_L/2)}{\widetilde\omega} \sim \frac{\pi^2}{2} \delta(\widetilde\omega) \ln\frac1{\widetilde{\lambda}_ L} 
	\end{split}
	\end{equation}
	for small but nonzero $\widetilde\lambda_L$. We note that the two other contributions to $K_0$, namely 
	$\partial_{\eta}\left[s_\eta \widetilde{K}_1\right]$ and  $\partial_{\eta}\left[s_\eta \widetilde{K}_2\right]$ have a finite $\eta\to 0$ limit.
	Therefore we have to leading order
	\begin{align}
	& \widetilde{K}_{\eta \to 0}=\lim_
	{\eta\to 0} \left(4\widetilde{K}_3 \widetilde{K}_4 \widetilde{K}_3\right)\sim \pi \ln (1/{\widetilde\lambda_L} )\,   
	\widetilde{P}  \\
	& \text{where} \,\,\;\; (\widetilde{P} f)( \widetilde{t}) \equiv \sech\frac{\widetilde{t}}2  \int \sech\frac{\widetilde{s}}2 \, f(\widetilde{s}) \mathrm{d} \widetilde{s} \,. \nonumber \label{eq:projector}
	\end{align}
	$\widetilde{P}$ can be readily diagonalized: its dominant eigenstate is $\propto \sech(\widetilde{t}/2)$, with eigenvalue $4$. Therefore, 
	\begin{equation}
	\widetilde{k}_{\eta \to 0} \sim 4\pi \ln (1/\widetilde{\lambda}_ L)  \,,
	\end{equation}
	for small $\widetilde\lambda_L$. Combining this with \eqref{eq:lambda_L:margimal}, we finally obtain the leading behavior of the Lyapunov exponent at small coupling constant
	\begin{equation}
	\lambda_L \sim \left(g\ln (1/g)\right) 2\pi T \label{eq:marginal_result} \,.
	\end{equation} 
	Here $g \equiv {4\lmax\gamma}/{\widehat\kappa} $ (in restored units) is the dimensionless coupling constant.
	The dependence on the coupling constant in the marginal Fermi liquid is distinct from both the Fermi liquid, for which it is linear in $g$ and the non-Fermi liquid, for which we obtained $\lambda_L$ independent of $g$ in the low temperature limit.

	\section{Discussion}\label{sec:conclusion}
	We have calculated the low-temperature behavior of the Lyapunov exponent $\lambda_L$ in a few large-$N$ systems, which are not fast scramblers, and whose low-temperature state is captured by perturbing a non-interacting limit. We stress again that the perturbation theory is applied to find the propagators, whereas the calculation of $\lambda_L$ involves a non-perturbative resummation of an infinity of ladder diagrams. Our analysis shows that the key factor that determines the behavior of $\lambda_L$ is how the thermal quasiparticle decay rate $1/\tau$ compares to temperature $T$ as both approach zero. When $1/\tau$ is much smaller, one has a Fermi liquid. The temperature dependence of $\lambda_L$ is dictated by $1/\tau$. It is also perturbative, vanishing with the interacting strength. If $1/\tau$ is much larger, one has a non-Fermi liquid, and $\lambda_L$ is $T$-linear, with a prefactor that remains constant even as the coupling constant approaches $0$. We expect the above results to apply generally despite being derived in a particular model, because  our analysis depends only on the scaling with temperature, and not the details of the ladder diagrams. In the case of the marginal Fermi liquid in low rank SYK, we found a non-analytical dependence on the coupling constant of the $T$-linear prefactor~\eqref{eq:marginal_result}. While this is a natural result that nicely interpolates the Fermi and non-Fermi liquid ones, its derivation did rely on specifics of our model. It will be interesting to see whether Ans\"atze like $\lambda_L \sim T / |\ln T|$, which is \textit{a priori} reasonable, can appear in other marginal Fermi liquids. 
	
	We close with two interesting observations. In the non-Fermi liquid states, the Lyapunov exponent can be arbitrary close the fast-scrambler value $\lambda_L = 2\pi T$, even if it is obtained in a perturbation theory. Now, it is known that fast scramblers like SYK have other outstanding properties, such as coherent scrambling~\cite{kitaev:soft} and enhanced teleportation capacity even at low temperatures ~\cite{Gao:2016bin,Maldacena:2017axo,Gao:2019nyj,Maldacena:2018lmt}. An interesting question is to what extent  ``second-class'' scramblers with $\lambda_L = c T < 2\pi T$ can facilitate teleportation at low temperatures.
	
	For the slow-scrambling Fermi liquids, we have been unable to show $\lambda_L > 0$ without explicitly computing it. Thus the question remains if there are examples of interacting large-$N$ model for which $\lambda$ is identically zero at non vanishing temperatures.
	(The low-rank SYK model with $\gamma = 0$ is for this purpose non-interacting~\cite{Bi:2017yvx,Kim:2019lwh}). It will be interesting to find either a general argument, or a counter example of non-scrambling system with $\lambda_L = 0$. 
	\begin{acknowledgments}
		We acknowledge support from the DOE grant DE-SC0019380, and from Gordon and Betty Moore Foundation's EPIC initiative, Grant GBMF4545 (X.C. and  E.A.).
	\end{acknowledgments}

	\bibliography{ref}

\begin{thebibliography}{26}%
\makeatletter
\providecommand \@ifxundefined [1]{%
 \@ifx{#1\undefined}
}%
\providecommand \@ifnum [1]{%
 \ifnum #1\expandafter \@firstoftwo
 \else \expandafter \@secondoftwo
 \fi
}%
\providecommand \@ifx [1]{%
 \ifx #1\expandafter \@firstoftwo
 \else \expandafter \@secondoftwo
 \fi
}%
\providecommand \natexlab [1]{#1}%
\providecommand \enquote  [1]{``#1''}%
\providecommand \bibnamefont  [1]{#1}%
\providecommand \bibfnamefont [1]{#1}%
\providecommand \citenamefont [1]{#1}%
\providecommand \href@noop [0]{\@secondoftwo}%
\providecommand \href [0]{\begingroup \@sanitize@url \@href}%
\providecommand \@href[1]{\@@startlink{#1}\@@href}%
\providecommand \@@href[1]{\endgroup#1\@@endlink}%
\providecommand \@sanitize@url [0]{\catcode `\\12\catcode `\$12\catcode
  `\&12\catcode `\#12\catcode `\^12\catcode `\_12\catcode `\%12\relax}%
\providecommand \@@startlink[1]{}%
\providecommand \@@endlink[0]{}%
\providecommand \url  [0]{\begingroup\@sanitize@url \@url }%
\providecommand \@url [1]{\endgroup\@href {#1}{\urlprefix }}%
\providecommand \urlprefix  [0]{URL }%
\providecommand \Eprint [0]{\href }%
\providecommand \doibase [0]{http://dx.doi.org/}%
\providecommand \selectlanguage [0]{\@gobble}%
\providecommand \bibinfo  [0]{\@secondoftwo}%
\providecommand \bibfield  [0]{\@secondoftwo}%
\providecommand \translation [1]{[#1]}%
\providecommand \BibitemOpen [0]{}%
\providecommand \bibitemStop [0]{}%
\providecommand \bibitemNoStop [0]{.\EOS\space}%
\providecommand \EOS [0]{\spacefactor3000\relax}%
\providecommand \BibitemShut  [1]{\csname bibitem#1\endcsname}%
\let\auto@bib@innerbib\@empty
\bibitem [{\citenamefont {{Larkin}}\ and\ \citenamefont
  {{Ovchinnikov}}(1969)}]{larkin}%
  \BibitemOpen
  \bibfield  {author} {\bibinfo {author} {\bibfnamefont {A.~I.}\ \bibnamefont
  {{Larkin}}}\ and\ \bibinfo {author} {\bibfnamefont {Yu.~N.}\ \bibnamefont
  {{Ovchinnikov}}},\ }\bibfield  {title} {\enquote {\bibinfo {title}
  {{Quasiclassical Method in the Theory of Superconductivity}},}\ }\href@noop
  {} {\bibfield  {journal} {\bibinfo  {journal} {Soviet Journal of Experimental
  and Theoretical Physics}\ }\textbf {\bibinfo {volume} {28}},\ \bibinfo
  {pages} {1200} (\bibinfo {year} {1969})}\BibitemShut {NoStop}%
\bibitem [{\citenamefont {Maldacena}\ \emph {et~al.}(2016)\citenamefont
  {Maldacena}, \citenamefont {Shenker},\ and\ \citenamefont
  {Stanford}}]{Maldacena:2015waa}%
  \BibitemOpen
  \bibfield  {author} {\bibinfo {author} {\bibfnamefont {Juan}\ \bibnamefont
  {Maldacena}}, \bibinfo {author} {\bibfnamefont {Stephen~H.}\ \bibnamefont
  {Shenker}}, \ and\ \bibinfo {author} {\bibfnamefont {Douglas}\ \bibnamefont
  {Stanford}},\ }\bibfield  {title} {\enquote {\bibinfo {title} {{A bound on
  chaos}},}\ }\href {\doibase 10.1007/JHEP08(2016)106} {\bibfield  {journal}
  {\bibinfo  {journal} {JHEP}\ }\textbf {\bibinfo {volume} {08}},\ \bibinfo
  {pages} {106} (\bibinfo {year} {2016})},\ \Eprint
  {http://arxiv.org/abs/1503.01409} {arXiv:1503.01409 [hep-th]} \BibitemShut
  {NoStop}%
\bibitem [{\citenamefont {Sachdev}\ and\ \citenamefont
  {Ye}(1993)}]{Sachdev:1992fk}%
  \BibitemOpen
  \bibfield  {author} {\bibinfo {author} {\bibfnamefont {Subir}\ \bibnamefont
  {Sachdev}}\ and\ \bibinfo {author} {\bibfnamefont {Jinwu}\ \bibnamefont
  {Ye}},\ }\bibfield  {title} {\enquote {\bibinfo {title} {{Gapless spin fluid
  ground state in a random, quantum Heisenberg magnet}},}\ }\href {\doibase
  10.1103/PhysRevLett.70.3339} {\bibfield  {journal} {\bibinfo  {journal}
  {Phys. Rev. Lett.}\ }\textbf {\bibinfo {volume} {70}},\ \bibinfo {pages}
  {3339} (\bibinfo {year} {1993})},\ \Eprint
  {http://arxiv.org/abs/cond-mat/9212030} {arXiv:cond-mat/9212030 [cond-mat]}
  \BibitemShut {NoStop}%
\bibitem [{\citenamefont {Georges}\ \emph {et~al.}(2001)\citenamefont
  {Georges}, \citenamefont {Parcollet},\ and\ \citenamefont
  {Sachdev}}]{george-parcollet}%
  \BibitemOpen
  \bibfield  {author} {\bibinfo {author} {\bibfnamefont {A.}~\bibnamefont
  {Georges}}, \bibinfo {author} {\bibfnamefont {O.}~\bibnamefont {Parcollet}},
  \ and\ \bibinfo {author} {\bibfnamefont {S.}~\bibnamefont {Sachdev}},\
  }\bibfield  {title} {\enquote {\bibinfo {title} {Quantum fluctuations of a
  nearly critical heisenberg spin glass},}\ }\href {\doibase
  10.1103/PhysRevB.63.134406} {\bibfield  {journal} {\bibinfo  {journal} {Phys.
  Rev. B}\ }\textbf {\bibinfo {volume} {63}},\ \bibinfo {pages} {134406}
  (\bibinfo {year} {2001})}\BibitemShut {NoStop}%
\bibitem [{\citenamefont {Kitaev}()}]{Kitaev:2015}%
  \BibitemOpen
  \bibfield  {author} {\bibinfo {author} {\bibfnamefont {Alexei}\ \bibnamefont
  {Kitaev}},\ }\bibfield  {title} {\enquote {\bibinfo {title} {{A simple model
  of quantum holography}},}\ }\href@noop {} {\ }\bibinfo {note}
  {\url{http://online.kitp.ucsb.edu/online/entangled15/kitaev/},\url{http://online.kitp.ucsb.edu/online/entangled15/kitaev2/}.
  Talks at KITP, April 7, 2015 and May 27, 2015}\BibitemShut {NoStop}%
\bibitem [{\citenamefont {Maldacena}\ and\ \citenamefont
  {Stanford}(2016)}]{syk}%
  \BibitemOpen
  \bibfield  {author} {\bibinfo {author} {\bibfnamefont {Juan}\ \bibnamefont
  {Maldacena}}\ and\ \bibinfo {author} {\bibfnamefont {Douglas}\ \bibnamefont
  {Stanford}},\ }\bibfield  {title} {\enquote {\bibinfo {title} {Remarks on the
  sachdev-ye-kitaev model},}\ }\href {\doibase 10.1103/PhysRevD.94.106002}
  {\bibfield  {journal} {\bibinfo  {journal} {Phys. Rev. D}\ }\textbf {\bibinfo
  {volume} {94}},\ \bibinfo {pages} {106002} (\bibinfo {year}
  {2016})}\BibitemShut {NoStop}%
\bibitem [{\citenamefont {Kitaev}\ and\ \citenamefont
  {Suh}(2018)}]{kitaev:soft}%
  \BibitemOpen
  \bibfield  {author} {\bibinfo {author} {\bibfnamefont {Alexei}\ \bibnamefont
  {Kitaev}}\ and\ \bibinfo {author} {\bibfnamefont {S.~Josephine}\ \bibnamefont
  {Suh}},\ }\bibfield  {title} {\enquote {\bibinfo {title} {The soft mode in
  the sachdev-ye-kitaev model and its gravity dual},}\ }\href@noop {}
  {\bibfield  {journal} {\bibinfo  {journal} {Journal of High Energy Physics}\
  }\textbf {\bibinfo {volume} {2018}},\ \bibinfo {pages} {183} (\bibinfo {year}
  {2018})}\BibitemShut {NoStop}%
\bibitem [{\citenamefont {Patel}\ and\ \citenamefont
  {Sachdev}(2017)}]{Patel1844}%
  \BibitemOpen
  \bibfield  {author} {\bibinfo {author} {\bibfnamefont {Aavishkar~A.}\
  \bibnamefont {Patel}}\ and\ \bibinfo {author} {\bibfnamefont {Subir}\
  \bibnamefont {Sachdev}},\ }\bibfield  {title} {\enquote {\bibinfo {title}
  {Quantum chaos on a critical fermi surface},}\ }\href {\doibase
  10.1073/pnas.1618185114} {\bibfield  {journal} {\bibinfo  {journal}
  {Proceedings of the National Academy of Sciences}\ }\textbf {\bibinfo
  {volume} {114}},\ \bibinfo {pages} {1844--1849} (\bibinfo {year} {2017})},\
  \Eprint
  {http://arxiv.org/abs/https://www.pnas.org/content/114/8/1844.full.pdf}
  {https://www.pnas.org/content/114/8/1844.full.pdf} \BibitemShut {NoStop}%
\bibitem [{\citenamefont {Banerjee}\ and\ \citenamefont
  {Altman}(2017)}]{sumilan}%
  \BibitemOpen
  \bibfield  {author} {\bibinfo {author} {\bibfnamefont {Sumilan}\ \bibnamefont
  {Banerjee}}\ and\ \bibinfo {author} {\bibfnamefont {Ehud}\ \bibnamefont
  {Altman}},\ }\bibfield  {title} {\enquote {\bibinfo {title} {Solvable model
  for a dynamical quantum phase transition from fast to slow scrambling},}\
  }\href {\doibase 10.1103/PhysRevB.95.134302} {\bibfield  {journal} {\bibinfo
  {journal} {Phys. Rev. B}\ }\textbf {\bibinfo {volume} {95}},\ \bibinfo
  {pages} {134302} (\bibinfo {year} {2017})}\BibitemShut {NoStop}%
\bibitem [{\citenamefont {Guo}\ \emph {et~al.}(2019)\citenamefont {Guo},
  \citenamefont {Gu},\ and\ \citenamefont {Sachdev}}]{guo}%
  \BibitemOpen
  \bibfield  {author} {\bibinfo {author} {\bibfnamefont {Haoyu}\ \bibnamefont
  {Guo}}, \bibinfo {author} {\bibfnamefont {Yingfei}\ \bibnamefont {Gu}}, \
  and\ \bibinfo {author} {\bibfnamefont {Subir}\ \bibnamefont {Sachdev}},\
  }\bibfield  {title} {\enquote {\bibinfo {title} {Transport and chaos in
  lattice sachdev-ye-kitaev models},}\ }\href {\doibase
  10.1103/PhysRevB.100.045140} {\bibfield  {journal} {\bibinfo  {journal}
  {Phys. Rev. B}\ }\textbf {\bibinfo {volume} {100}},\ \bibinfo {pages}
  {045140} (\bibinfo {year} {2019})}\BibitemShut {NoStop}%
\bibitem [{\citenamefont {Patel}\ \emph {et~al.}(2017)\citenamefont {Patel},
  \citenamefont {Chowdhury}, \citenamefont {Sachdev},\ and\ \citenamefont
  {Swingle}}]{diffusive}%
  \BibitemOpen
  \bibfield  {author} {\bibinfo {author} {\bibfnamefont {Aavishkar~A.}\
  \bibnamefont {Patel}}, \bibinfo {author} {\bibfnamefont {Debanjan}\
  \bibnamefont {Chowdhury}}, \bibinfo {author} {\bibfnamefont {Subir}\
  \bibnamefont {Sachdev}}, \ and\ \bibinfo {author} {\bibfnamefont {Brian}\
  \bibnamefont {Swingle}},\ }\bibfield  {title} {\enquote {\bibinfo {title}
  {Quantum butterfly effect in weakly interacting diffusive metals},}\ }\href
  {\doibase 10.1103/PhysRevX.7.031047} {\bibfield  {journal} {\bibinfo
  {journal} {Phys. Rev. X}\ }\textbf {\bibinfo {volume} {7}},\ \bibinfo {pages}
  {031047} (\bibinfo {year} {2017})}\BibitemShut {NoStop}%
\bibitem [{\citenamefont {Liao}\ and\ \citenamefont
  {Galitski}(2018)}]{Galitski}%
  \BibitemOpen
  \bibfield  {author} {\bibinfo {author} {\bibfnamefont {Yunxiang}\
  \bibnamefont {Liao}}\ and\ \bibinfo {author} {\bibfnamefont {Victor}\
  \bibnamefont {Galitski}},\ }\bibfield  {title} {\enquote {\bibinfo {title}
  {Nonlinear sigma model approach to many-body quantum chaos: Regularized and
  unregularized out-of-time-ordered correlators},}\ }\href {\doibase
  10.1103/PhysRevB.98.205124} {\bibfield  {journal} {\bibinfo  {journal} {Phys.
  Rev. B}\ }\textbf {\bibinfo {volume} {98}},\ \bibinfo {pages} {205124}
  (\bibinfo {year} {2018})}\BibitemShut {NoStop}%
\bibitem [{\citenamefont {Chowdhury}\ and\ \citenamefont
  {Swingle}(2017)}]{PhysRevD.96.065005}%
  \BibitemOpen
  \bibfield  {author} {\bibinfo {author} {\bibfnamefont {Debanjan}\
  \bibnamefont {Chowdhury}}\ and\ \bibinfo {author} {\bibfnamefont {Brian}\
  \bibnamefont {Swingle}},\ }\bibfield  {title} {\enquote {\bibinfo {title}
  {Onset of many-body chaos in the $o(n)$ model},}\ }\href {\doibase
  10.1103/PhysRevD.96.065005} {\bibfield  {journal} {\bibinfo  {journal} {Phys.
  Rev. D}\ }\textbf {\bibinfo {volume} {96}},\ \bibinfo {pages} {065005}
  (\bibinfo {year} {2017})}\BibitemShut {NoStop}%
\bibitem [{\citenamefont {Danshita}\ \emph {et~al.}(2017)\citenamefont
  {Danshita}, \citenamefont {Hanada},\ and\ \citenamefont {Tezuka}}]{masaki16}%
  \BibitemOpen
  \bibfield  {author} {\bibinfo {author} {\bibfnamefont {Ippei}\ \bibnamefont
  {Danshita}}, \bibinfo {author} {\bibfnamefont {Masanori}\ \bibnamefont
  {Hanada}}, \ and\ \bibinfo {author} {\bibfnamefont {Masaki}\ \bibnamefont
  {Tezuka}},\ }\bibfield  {title} {\enquote {\bibinfo {title} {{Creating and
  probing the Sachdev–Ye–Kitaev model with ultracold gases: Towards
  experimental studies of quantum gravity}},}\ }\href {\doibase
  10.1093/ptep/ptx108} {\bibfield  {journal} {\bibinfo  {journal} {Progress of
  Theoretical and Experimental Physics}\ }\textbf {\bibinfo {volume} {2017}}
  (\bibinfo {year} {2017}),\ 10.1093/ptep/ptx108}\BibitemShut {NoStop}%
\bibitem [{\citenamefont {Bi}\ \emph {et~al.}(2017)\citenamefont {Bi},
  \citenamefont {Jian}, \citenamefont {You}, \citenamefont {Pawlak},\ and\
  \citenamefont {Xu}}]{Bi:2017yvx}%
  \BibitemOpen
  \bibfield  {author} {\bibinfo {author} {\bibfnamefont {Zhen}\ \bibnamefont
  {Bi}}, \bibinfo {author} {\bibfnamefont {Chao-Ming}\ \bibnamefont {Jian}},
  \bibinfo {author} {\bibfnamefont {Yi-Zhuang}\ \bibnamefont {You}}, \bibinfo
  {author} {\bibfnamefont {Kelly~Ann}\ \bibnamefont {Pawlak}}, \ and\ \bibinfo
  {author} {\bibfnamefont {Cenke}\ \bibnamefont {Xu}},\ }\bibfield  {title}
  {\enquote {\bibinfo {title} {{Instability of the non-Fermi liquid state of
  the Sachdev-Ye-Kitaev Model}},}\ }\href {\doibase 10.1103/PhysRevB.95.205105}
  {\bibfield  {journal} {\bibinfo  {journal} {Phys. Rev.}\ }\textbf {\bibinfo
  {volume} {B95}},\ \bibinfo {pages} {205105} (\bibinfo {year} {2017})},\
  \Eprint {http://arxiv.org/abs/1701.07081} {arXiv:1701.07081
  [cond-mat.str-el]} \BibitemShut {NoStop}%
\bibitem [{\citenamefont {Wang}(2020)}]{wang}%
  \BibitemOpen
  \bibfield  {author} {\bibinfo {author} {\bibfnamefont {Yuxuan}\ \bibnamefont
  {Wang}},\ }\bibfield  {title} {\enquote {\bibinfo {title} {Solvable
  strong-coupling quantum-dot model with a non-fermi-liquid pairing
  transition},}\ }\href {\doibase 10.1103/PhysRevLett.124.017002} {\bibfield
  {journal} {\bibinfo  {journal} {Phys. Rev. Lett.}\ }\textbf {\bibinfo
  {volume} {124}},\ \bibinfo {pages} {017002} (\bibinfo {year}
  {2020})}\BibitemShut {NoStop}%
\bibitem [{\citenamefont {Esterlis}\ and\ \citenamefont
  {Schmalian}(2019)}]{esterlis}%
  \BibitemOpen
  \bibfield  {author} {\bibinfo {author} {\bibfnamefont {Ilya}\ \bibnamefont
  {Esterlis}}\ and\ \bibinfo {author} {\bibfnamefont {J\"org}\ \bibnamefont
  {Schmalian}},\ }\bibfield  {title} {\enquote {\bibinfo {title} {Cooper
  pairing of incoherent electrons: An electron-phonon version of the
  sachdev-ye-kitaev model},}\ }\href {\doibase 10.1103/PhysRevB.100.115132}
  {\bibfield  {journal} {\bibinfo  {journal} {Phys. Rev. B}\ }\textbf {\bibinfo
  {volume} {100}},\ \bibinfo {pages} {115132} (\bibinfo {year}
  {2019})}\BibitemShut {NoStop}%
\bibitem [{\citenamefont {Kim}\ \emph {et~al.}(2020)\citenamefont {Kim},
  \citenamefont {Cao},\ and\ \citenamefont {Altman}}]{Kim:2019lwh}%
  \BibitemOpen
  \bibfield  {author} {\bibinfo {author} {\bibfnamefont {Jaewon}\ \bibnamefont
  {Kim}}, \bibinfo {author} {\bibfnamefont {Xiangyu}\ \bibnamefont {Cao}}, \
  and\ \bibinfo {author} {\bibfnamefont {Ehud}\ \bibnamefont {Altman}},\
  }\bibfield  {title} {\enquote {\bibinfo {title} {{Low-rank Sachdev-Ye-Kitaev
  models}},}\ }\href {\doibase 10.1103/PhysRevB.101.125112} {\bibfield
  {journal} {\bibinfo  {journal} {Phys. Rev. B}\ }\textbf {\bibinfo {volume}
  {101}},\ \bibinfo {pages} {125112} (\bibinfo {year} {2020})},\ \Eprint
  {http://arxiv.org/abs/1910.10173} {arXiv:1910.10173 [cond-mat.str-el]}
  \BibitemShut {NoStop}%
\bibitem [{\citenamefont {Kim}\ and\ \citenamefont {Cao}(2020)}]{Kim:2020mho}%
  \BibitemOpen
  \bibfield  {author} {\bibinfo {author} {\bibfnamefont {Jaewon}\ \bibnamefont
  {Kim}}\ and\ \bibinfo {author} {\bibfnamefont {Xiangyu}\ \bibnamefont
  {Cao}},\ }\bibfield  {title} {\enquote {\bibinfo {title} {{Comment on
  "Chaotic-Integrable Transition in the Sachdev-Ye-Kitaev Model"}},}\
  }\href@noop {} {\  (\bibinfo {year} {2020})},\ \Eprint
  {http://arxiv.org/abs/2004.05313} {arXiv:2004.05313 [cond-mat.stat-mech]}
  \BibitemShut {NoStop}%
\bibitem [{\citenamefont {Kobrin}\ \emph {et~al.}(2020)\citenamefont {Kobrin},
  \citenamefont {Yang}, \citenamefont {Kahanamoku-Meyer}, \citenamefont
  {Olund}, \citenamefont {Moore}, \citenamefont {Stanford},\ and\ \citenamefont
  {Yao}}]{kobrin2020many}%
  \BibitemOpen
  \bibfield  {author} {\bibinfo {author} {\bibfnamefont {Bryce}\ \bibnamefont
  {Kobrin}}, \bibinfo {author} {\bibfnamefont {Zhenbin}\ \bibnamefont {Yang}},
  \bibinfo {author} {\bibfnamefont {Gregory~D}\ \bibnamefont
  {Kahanamoku-Meyer}}, \bibinfo {author} {\bibfnamefont {Christopher~T}\
  \bibnamefont {Olund}}, \bibinfo {author} {\bibfnamefont {Joel~E}\
  \bibnamefont {Moore}}, \bibinfo {author} {\bibfnamefont {Douglas}\
  \bibnamefont {Stanford}}, \ and\ \bibinfo {author} {\bibfnamefont {Norman~Y}\
  \bibnamefont {Yao}},\ }\bibfield  {title} {\enquote {\bibinfo {title}
  {Many-body chaos in the sachdev-ye-kitaev model},}\ }\href@noop {} {\bibfield
   {journal} {\bibinfo  {journal} {arXiv:2002.05725}\ } (\bibinfo {year}
  {2020})}\BibitemShut {NoStop}%
\bibitem [{Note1()}]{Note1}%
  \BibitemOpen
  \bibinfo {note} {For any $\lambda _L$, this space is closed under the action
  of $K$, so long as $K(t_1, \protect \dots , t_4) =K(t_1 + \delta t, \protect
  \dots , t_4+\delta t) $ is invariant under time translation.}\BibitemShut
  {Stop}%
\bibitem [{Note2()}]{Note2}%
  \BibitemOpen
  \bibinfo {note} {We have omitted the boson kinetic term, which would be
  irrelevant in any case~\cite {Kim:2019lwh}.}\BibitemShut {Stop}%
\bibitem [{\citenamefont {Gao}\ \emph {et~al.}(2017)\citenamefont {Gao},
  \citenamefont {Jafferis},\ and\ \citenamefont {Wall}}]{Gao:2016bin}%
  \BibitemOpen
  \bibfield  {author} {\bibinfo {author} {\bibfnamefont {Ping}\ \bibnamefont
  {Gao}}, \bibinfo {author} {\bibfnamefont {Daniel~Louis}\ \bibnamefont
  {Jafferis}}, \ and\ \bibinfo {author} {\bibfnamefont {Aron~C.}\ \bibnamefont
  {Wall}},\ }\bibfield  {title} {\enquote {\bibinfo {title} {{Traversable
  Wormholes via a Double Trace Deformation}},}\ }\href {\doibase
  10.1007/JHEP12(2017)151} {\bibfield  {journal} {\bibinfo  {journal} {JHEP}\
  }\textbf {\bibinfo {volume} {12}},\ \bibinfo {pages} {151} (\bibinfo {year}
  {2017})},\ \Eprint {http://arxiv.org/abs/1608.05687} {arXiv:1608.05687
  [hep-th]} \BibitemShut {NoStop}%
\bibitem [{\citenamefont {Maldacena}\ \emph {et~al.}(2017)\citenamefont
  {Maldacena}, \citenamefont {Stanford},\ and\ \citenamefont
  {Yang}}]{Maldacena:2017axo}%
  \BibitemOpen
  \bibfield  {author} {\bibinfo {author} {\bibfnamefont {Juan}\ \bibnamefont
  {Maldacena}}, \bibinfo {author} {\bibfnamefont {Douglas}\ \bibnamefont
  {Stanford}}, \ and\ \bibinfo {author} {\bibfnamefont {Zhenbin}\ \bibnamefont
  {Yang}},\ }\bibfield  {title} {\enquote {\bibinfo {title} {{Diving into
  traversable wormholes}},}\ }\href {\doibase 10.1002/prop.201700034}
  {\bibfield  {journal} {\bibinfo  {journal} {Fortsch. Phys.}\ }\textbf
  {\bibinfo {volume} {65}},\ \bibinfo {pages} {1700034} (\bibinfo {year}
  {2017})},\ \Eprint {http://arxiv.org/abs/1704.05333} {arXiv:1704.05333
  [hep-th]} \BibitemShut {NoStop}%
\bibitem [{\citenamefont {Gao}\ and\ \citenamefont
  {Jafferis}(2019)}]{Gao:2019nyj}%
  \BibitemOpen
  \bibfield  {author} {\bibinfo {author} {\bibfnamefont {Ping}\ \bibnamefont
  {Gao}}\ and\ \bibinfo {author} {\bibfnamefont {Daniel~Louis}\ \bibnamefont
  {Jafferis}},\ }\bibfield  {title} {\enquote {\bibinfo {title} {{A Traversable
  Wormhole Teleportation Protocol in the SYK Model}},}\ }\href@noop {} {\
  (\bibinfo {year} {2019})},\ \Eprint {http://arxiv.org/abs/1911.07416}
  {arXiv:1911.07416} \BibitemShut {NoStop}%
\bibitem [{\citenamefont {Maldacena}\ and\ \citenamefont
  {Qi}(2018)}]{Maldacena:2018lmt}%
  \BibitemOpen
  \bibfield  {author} {\bibinfo {author} {\bibfnamefont {Juan}\ \bibnamefont
  {Maldacena}}\ and\ \bibinfo {author} {\bibfnamefont {Xiao-Liang}\
  \bibnamefont {Qi}},\ }\bibfield  {title} {\enquote {\bibinfo {title}
  {{Eternal traversable wormhole}},}\ }\href@noop {} {\  (\bibinfo {year}
  {2018})},\ \Eprint {http://arxiv.org/abs/1804.00491} {arXiv:1804.00491}
  \BibitemShut {NoStop}%
\end{thebibliography}%
	
\end{document}